\documentclass[useAMS,usenatbib]{mn2e}

\usepackage{natbib}
\usepackage{graphicx}
\usepackage{listings}
\usepackage[intlimits]{amsmath}%option changes placement of integral limits
\usepackage{txfonts}
\usepackage{bm}
\usepackage{amssymb}
\usepackage{hyperref}
\usepackage{amsmath}
\usepackage{tabularx} 
\usepackage{color}
\usepackage{comment}

\newcommand{\beq}{\begin{equation}}
\newcommand{\eeq}{\end{equation}}

\def\be{\begin{equation}}
\def\ee{\end{equation}}
\def\ba{\begin{eqnarray}}
\def\ea{\end{eqnarray}}
\def\go{\mathrel{\raise.3ex\hbox{$>$}\mkern-14mu
             \lower0.6ex\hbox{$\sim$}}}
\def\lo{\mathrel{\raise.3ex\hbox{$<$}\mkern-14mu
             \lower0.6ex\hbox{$\sim$}}}

\def\apj{Astrophys. J.}
\def\apjl{Astrophys. J. Lett.}

\def\aap{Astron. Astrophys. }

\def\araa{Ann.\ Rev. Astron. Astroph. }

\def\mnras{Mon. Not. Roy. Astron. Soc. }

\title[Plastic Flows]{Axisymmetric magneto-plastic evolution of neutron-star crusts}
\author[K.N. Gourgouliatos \& S.K. Lander]{{Konstantinos N. Gourgouliatos $^{1}$}\thanks{kngourg@upatras.gr} \& {Samuel K. Lander  $^{2}$}\\
\parbox{\textwidth}{$^{1}$University of Patras, Department of Physics, 26504, Patras, Greece }\\
\parbox{\textwidth}{$^{2}$School of Physics, University of East Anglia, Norwich NR4 7TJ, UK} }

\begin{document}

\date{Accepted -. Received -; in original form -}
\pagerange{\pageref{firstpage}--\pageref{lastpage}} \pubyear{-}
\maketitle

\label{firstpage}

\begin{abstract}
Magnetic field evolution in neutron-star crusts is driven by the Hall effect and Ohmic dissipation, for as long as the crust is sufficiently strong to absorb Maxwell stresses exerted by the field and thus make the momentum equation redundant. For the strongest neutron-star fields, however, stresses build to the point of  crustal failure, at which point the standard evolution equations are no longer valid. Here, we study the evolution of the magnetic field of the crust up to and beyond crustal failure, whence the crust begins to flow plastically. We perform global axisymmetric evolutions, exploring different types of failure affecting a limited region of the crust. We find that a plastic flow does not simply suppress the Hall effect even in the regime of a low plastic viscosity, but it rather leads to non-trivial evolution -- in some cases even overreacting and enhancing the impact of the Hall effect. Its impact is more pronouced in the toroidal field, with the differences on the poloidal field being less substantial. We argue that both the nature of magnetar bursts and their spindown evolution will be affected by plastic flow, so that observations of these phenomena may help to constrain the way the crust fails.
\end{abstract}


\section{Introduction}

The magnetic field in the crust of strongly magnetised neutron stars evolves due to the Hall effect, which is the advection of the magnetic flux by the electron fluid. This process dominates over Ohmic decay, which is caused by the finite conductivity of the crust, for magnetic fields above $10^{12}-10^{13}$G 
\citep{Goldreich:1992, Cumming:2004}. A key assumption in the Hall-Ohmic evolution scheme is that the forces acting on the crust are in balance and the system remains in equilibrium. This holds, provided that any deformation of the crust is below its elastic limit. At the upper end of neutron star magnetic fields, however, the magnetic field may become sufficiently intense to exceed the yield stress, at which point the crust will fail. Such crustal-failure events have been related to explosive activity in magnetars such as bursts and flares \citep{Thompson:1995, Turolla:2015,Kaspi:2017, Gourgouliatos:2018b, Esposito:2021}. In this context, the magnetic field of magnetars produces Maxwell stress leading to shear deformations close or beyond the elastic limit of the crust \citep{Horowitz:2009, Perna:2011, Pons:2011, Levin:2012, Lander:2016, Bransgrove:2018}. The outcome of these failures may lead to short-lived events of individual outbursts \citep{Beloborodov:2014, Li:2016, Beloborodov:2016}. 

While it is reasonable to expect that these explosive events are connected with the crust's elastic limit being exceeded, it is not clear how the magnetic field evolves beyond that point. The Hall-MHD evolution scheme is not valid, as the assumption of an elastically deformed crust does not hold any more. A likely scenario is that once the elastic limit is reached the crust enters a state of slow deformation or flow, where the Maxwell stress of the magnetic field drives the evolution of the crust. At the same time, the electrons continue to advect the magnetic field and the two effects act simultaneously.

In previous work \citep{Lander:2019} we considered the combination of Hall-MHD evolution and a plastic flow in a Cartesian domain, with plane-parallel symmetry. The Cartesian domain represented a slab of the neutron star crust that becomes stressed by the intense magnetic field, eventually exceeding the elastic limit and failing. We modelled the plastic flow that was initiated once the slab of crust had failed, assuming that the field then evolved due to the combination of plastic flow and the Hall effect in the entire domain. Although this study could not capture the global impact of this evolution on the crustal magnetic field, it allowed us to explore some of the characteristics of magneto-plastic flow and avoid the additional complexity associated with the possibility of having plastic flow in some regions of the crust and not others.

In the work reported here we extend our study to a global axisymmetric model simulating the magnetic field evolution in the entire crust. We simulate the magnetic field evolution due to the Hall-Ohmic effect and account for the plastic flow if the crust fails. Our work complements and extends the recent study of \cite{Kojima:2020}, who studied global crustal field evolution under a viscous flow, using the formalism of \cite{Lander:2016, Lander:2019}. It was found that for a strong magnetic ($B>10^{14}$ G) field and low plastic viscosity $10^{36}-10^{37}$ g cm$^{-1}$s$^{-1}$ a significant fraction of the energy is transferred into the bulk flow energy. Furthermore, these effects can lead to the magnetic deformation of the crust \citep{Kojima:2021}.

The work of \cite{Kojima:2020, Kojima:2021} makes, however, two key simplifications that deserve further study. Firstly, they start their simulations assuming the crust is already in a plastic regime, avoiding the challenging issue of diagnosing and applying an elastic failure criterion in their simulations. Secondly, by simulating a viscous crust they effectively assume that any crustal failure is global, so that the entire crust yields together, whether or not the local value of the stress is always very high. If, instead, the plastic flow is confined to the region where the yield stress is exceeded, or perhaps also its environs, the evolutionary path taken by the magnetic field may be very different. In this context, the profiles of magnetar outbursts suggest that the emitting region is not the entire crust but a limited fraction of it \citep{Tiengo:2008, Alford:2016, CotiZelati:2018}. This suggests that most of the crust remains intact despite the magnetar being in active phase. We further note that whilst we adopted a global flow in our previous study \citep{Lander:2019}, the simulations were intrinsically local, with the domain being a square slab with sides of length $0.5$ km. Thus, a major aim of this paper is to explore the impact of different types of failure and consequently plastic flow, addressing the differences between global and local flows. 

The plan of the paper is as follows. Section \ref{MATH} contains the setup of the problem, the relevant equations and the crust model. In section \ref{SIMULATIONS}, we discuss the numerical approach implemented for the integration of the differential equations of the system. In section \ref{RESULTS} we present the results of the simulations. We discuss their implications for strongly magnetised neutron stars in section \ref{DISCUSSION}. We conclude in section \ref{CONCLUSIONS}.

%%%%%%%%%%%%%%%%%%%%%%%%%%%%%%%%%%%%%%%%%%
\section{Problem setup}
\label{MATH}

\subsection{Magnetic induction equation}

Let us consider an axisymmetric magnetic field in spherical coordinates $(r,\theta,\phi)$
\beq
{\bf B}= B_{r}(r,\theta){\bf \hat{r}}+B_{\theta}(r,\theta)\bm {\hat{\theta}}+B_{\phi}(r,\theta)\bm{\hat{\phi}}\,.
\eeq
We assume that the evolution is driven by the flow of the electron fluid and the plastic deformation of the lattice. The former effect is approximated by the Hall-Ohmic evolution \citep{Goldreich:1992}, whereas the latter is described by a plastic flow. Taking these into account, the magnetic field induction equation becomes:
\beq
\partial_{t} {\bm B} = -\nabla\times \left[\left(\frac{c}{4 \pi e n_e}\left(\nabla \times {\bm B}\right)-{\bm v}_{pl}\right)\times {\bm B}  +\frac{c^2}{4 \pi \sigma}\nabla \times {\bm B}\right]\,,\
\label{HALL_EQ}
\eeq
where $n_e$ is the electron number density, $\sigma$ the electrical conductivity, $e$ is the elementary charge, ${\bm v}_{pl}$ the plastic flow velocity and $c$ the speed of light. 
The first term in the bracket results from the electron-fluid motion, the second one is the plastic flow velocity ${\bm v}_{pl}$ and the last one is due to the Ohmic dissipation. We can estimate the ratio of the Hall to the Ohmic term through the dimensionless Hall parameter:
\beq
R_{H}=\frac{\sigma |B|}{e c n_{e}}\,.
\label{RH}
\eeq
In the absence of a plastic flow, the Hall-Ohmic equation can be solved by direct numerical integration, as the knowledge of the microphysics determining $n_e$ and $\sigma$ suffice, in principle, for a numerical solution. We note, however, that the non-linear nature of the equation makes the solution of this equation far from trivial. 

Further complexity is added to the problem once the plastic flow is included. The details of how a neutron-star crust fails are not well understood - and so there is no longer an unambiguous route to understanding the crust's evolution in this case. In any case, an additional equation is needed to determine the plastic flow velocity that now appears in equation \eqref{HALL_EQ}. We follow the basic principles of the formulation of \cite{Lander:2019}, where we approximate the plastic flow as a Stokes flow. The Laplacian of the plastic flow velocity is equal to the divergence of the traceless part of the stress tensor of the current crust state, the stressed crust, minus the field arising from the unstressed crust, corresponding to the initial state, (we refer to this state as ``reference state" in \cite{Lander:2019}). This results in a Poisson partial differential equation 
\beq
4 \pi \nu \nabla^2 \bm{v}_{pl}=(\bm{B_0}\cdot \nabla)\bm{B_0}-(\bm{B}\cdot \nabla)\bm{B}-\frac{1}{3}\nabla(B_0^2-B^2) \, ,
\label{Laplacian}
\eeq
where a subscript $0$ denotes the magnetic field of the initial state. %We evaluate the terms of the above equation:
%
%\begin{eqnarray}
%&(\bm{B}\cdot \nabla)\bm{B}=\left(B_r \partial_r B_r+\frac{B_{\theta} \partial_{\theta} B_r}{r}-\frac{B_{\theta}^2}{r}-\frac{B_{\phi}^2 }{r}\right)\bm{\hat{r}}\nonumber\\
%&+\left(B_r\partial_r B_{\theta}+\frac{B_{\theta}\partial_{\theta}B_{\theta}}{r}+\frac{B_{\theta}B_{r}}{r}-\frac{\cos\theta  B_{\phi}^2}{ r\sin\theta} \right) \bm{\hat{\theta}}\nonumber \\
%&+\left(B_r\partial_r B_{\phi}+\frac{B_\theta\partial_{\theta}B_{\phi}}{r}+\frac{B_{\phi}B_{r}}{r}+\frac{\cos\theta B_{\phi}B_{\theta}}{r\sin\theta} \right)\bm{\hat{\phi}}\,.
%\label{BDB}
%\end{eqnarray}
%
%\begin{eqnarray}
%&\nabla B^2= 2\left(B_r \partial_r B_r+B_{\theta}\partial_r B_{\theta}+B_{\phi}\partial_r B_{\phi}\right)\bm{\hat{r}}\nonumber\\
%&+\frac{2}{r}\left(B_{r}\partial_{\theta}B_{r}+B_{\theta}\partial_{\theta}B_{\theta}+B_{\phi}\partial_{\theta}B_{\phi}\right)\bm{\hat{\theta}}\,.
%\label{DBB}
%\end{eqnarray}
%
We assume that any motion of the crust is incompressible. Thus, the continuity equation for the plastic flow velocity becomes
\beq
\nabla \cdot \bm{v}_{pl} = 0\,.
\eeq
Given that the crust is stably stratified in the radial direction, any radial displacement due to the plastic flow is zero, leading to a zero plastic flow radial velocity $v_{pl,r}=0$. Combining this with the continuity equation and taking into account axisymmetry, it reduces to the following differential equation 
\beq
\frac{1}{r\sin\theta}\partial_{\theta}\left(v_{pl,\theta}\sin\theta\right)=0\,,
\eeq
for which the only physically acceptable solution is $v_{pl,\theta}=0$. Therefore, the plastic flow is only along the azimuthal direction $\bm{v}_{pl}=v_{pl}\bm{\hat{\phi}}$. %Substituting this into the Laplacian of equation (\ref{Laplacian}) we obtain
%
%\begin{eqnarray}
%\nabla^2 \bm{v}_{pl}=\left(\partial_{rr}v_{pl}+\frac{1}{r^2}\partial_{\theta\theta}v_{pl}+\frac{2}{r}\partial_{r}v_{pl}+\frac{\cos\theta}{r^2\sin\theta}\partial_{\theta}v_{pl}\right. \nonumber \\
%\left.-\frac{1}{r^2\sin^2\theta}v_{pl}\right)\bm{\hat{\phi}}\,.
%\label{DDvpl}
%\end{eqnarray}
%
The only component of equation \eqref{Laplacian} that needs to be evaluated is the azimuthal one $\left(\bm{B} \cdot \nabla \right) \bm{B}|_{\phi}$ %which is then combined with the Laplacian from equation \eqref{DDvpl} and eventually 
and leads to determination of the flow velocity. We note that the magnetic pressure gradient term $\nabla B^2$%, equation \eqref{DBB}, 
does not have any component in the $\phi$ direction because of axisymmetry.   

In axisymmetry, we can express the magnetic field in terms of two scalar functions proportional to the poloidal flux $2\pi \Psi(r, \theta)$ and electric current $cI(r, \theta)/2$ that pass through a spherical cap centered on the axis of the system, at distance $r$ from the origin and semi-opening angle $\theta$. The magnetic field takes the following form:
\beq
{\bf B} =\nabla  \Psi(r,\theta) \times \nabla \phi + I(r, \theta) \nabla \phi\,.
\label{B_FIELD}
\eeq
Note that the toroidal magnetic field is expressed in terms of the poloidal current. The above expression satisfies by construction Gauss' law for the magnetic field $\nabla \cdot {\bf B}=0$. Substituting the expression of the magnetic field from equation \eqref{B_FIELD} into equation \eqref{HALL_EQ} we obtain two coupled partial differential equation for the scalars $\Psi$ and $I$. The first one for the poloidal field evolution is not directly affected by the plastic flow and is given by the following expression
\beq
\partial_t \Psi-r^2\sin^2\theta \chi \left(\nabla I \times \nabla \phi\right)\cdot \nabla \Psi =\frac{c^2}{4 \pi \sigma}\Delta^{\star} \Psi\,,
\label{dPsi}
\eeq
where we have set as in \cite{Reisenegger:2007}
\beq
\chi=\frac{c}{4\pi e n_e r^2\sin^2\theta}\,,
\eeq
and the Grad-Shafranov operator
\beq
\Delta^{\star}=\frac{\partial^2}{\partial r^2}+\frac{\sin \theta}{r^2}\frac{\partial}{\partial \theta}\left(\frac{1}{\sin\theta}\frac{\partial}{\partial \theta}\right)\, .
\label{GSop}
\eeq
The electron fluid angular velocity is 
\beq
\Omega_e=\chi \Delta^{\star}\Psi,
\label{Omega}
\eeq
and we also define the plastic flow angular velocity:
\beq
\Omega_{pl}=\frac{v_{pl}}{r\sin\theta}.
\eeq
The second equation is related to the  toroidal field evolution. In addition to the terms arising by the Hall effect, it is directly affected by the plastic flow velocity
\begin{eqnarray}
\frac{\partial I}{\partial t}+r^2\sin^2\theta\left\{\left[\nabla\left(\Omega_{e}+\Omega_{pl}\right)\times \nabla \phi\right]\cdot \nabla \Psi+I\left(\nabla \chi \times \nabla \phi\right)\cdot \nabla I\right\}\nonumber \\
=\frac{c^2}{4\pi \sigma}\left(\Delta^{\star} I-\frac{1}{\sigma}\nabla I \cdot \nabla \sigma\right)
\label{dI}
\end{eqnarray}
Thus, integration of equations (\ref{dPsi}),  (\ref{dI}) and use of equation (\ref{Laplacian}) with an appropriate failure criterion for the evaluation of the plastic flow velocity will allow the determination of the magnetic field evolution.

\subsection{Plastic flow initiation}
\label{PFI}

Directly before a neutron star's crust freezes, the stellar structure (including its magnetic field) is that of a fluid body, with no shear stresses. It follows that the crust forms in an unstressed state. In this work we consider only stresses that build up over time due to the evolving crustal magnetic field deviating from its initial state, although rotation will generally also add to the crust's stress.
For a sufficiently weak magnetic field, stresses will never grow enough to induce failure of the star's crust, and so the plastic flow velocity will remain identically zero. In this case, any deformation stays within the crust's elastic limit and the Hall-Ohmic evolution suffices for the description of the magnetic field evolution in the crust. For typical magnetar-strength fields, however, it is quite likely that Maxwell stresses will become strong enough to lead to crustal failure. It is not fully resolved, however, how these failures start and progress. 

In our approach, we use the modified von Mises criterion as described in \cite{Lander:2019}. In particular, the crust in our model fails if the following inequality is satisfied:
\beq
\tau_{el}\leq \frac{1}{4\pi} \sqrt{\frac{1}{3} B_0^4+\frac{1}{3}B^4+\frac{1}{3}B_0^2B^2-\left({\bm B}\cdot {\bm B_0}\right)^2}\,,
\label{FAILURE_EQ}
\eeq
where $\tau_{el}$ is the critical value of the stress the crust can support and is determined by the microphysics of the crust. 

Even if the crust fails somewhere, we further need to answer how such failures propagate in the crust and whether they lead to a flow localised only in the region where the inequality (\ref{FAILURE_EQ}) holds, or extend over larger parts of the crust. Given these uncertainties, we simulate three types of failure: a local, an intermediate and a global one, which correspond to different treatments of equation \ref{Laplacian}. 

 In the case of a local failure, we assume that the plastic flow velocity is given by the solution of equation (\ref{Laplacian}) in the region where the failure criterion is satisfied. Apart from this region, the plastic flow velocity is set to zero everywhere else in the crust. This way, the only part that flows is the region where the condition described in inequality (\ref{FAILURE_EQ}) is satisfied.

In the intermediate case,   we solve equation (\ref{Laplacian}) demanding that its right-hand-side is non-zero in the part of the crust where the failure criterion (\ref{FAILURE_EQ}) is satisfied, and zero elsewhere. Thus, the plastic flow velocity can be non-zero anywhere in the crust, however, a source term appears in equation (\ref{Laplacian}) only in the region where the failure criterion is satisfied. This provides a smoother transition of the plastic flow velocity between the regions of plastic flow and the rest of the star, whereas in the local flow, the transition between the failed region and the rest of the crust is rather sharp. 

Finally, in the scenario of a global failure, a plastic flow begins everywhere in the crust if the criterion of equation \eqref{FAILURE_EQ} is fulfilled even in a single point in the crust. The plastic flow velocity corresponds to the solution of equation \eqref{Laplacian} everywhere in the crust and thus is non-zero everywhere, although will be largest in the failed region. This last case is rather extreme and physically unlikely, but it is worth studying to set the maximum possible plastic flow that can result from our formalism. 

\begin{figure}
    \includegraphics[width=0.45\textwidth]{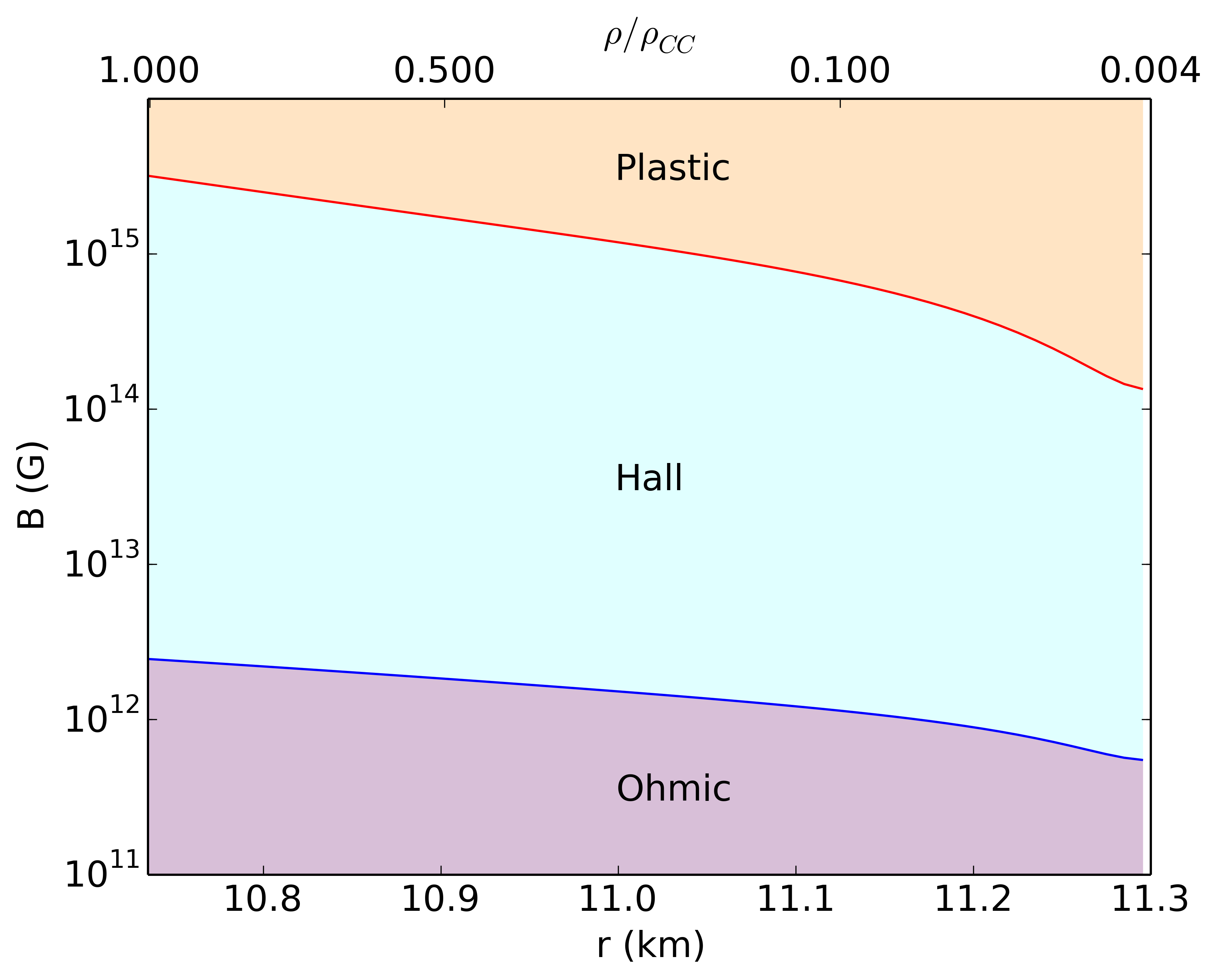}
    \caption{Dominance of Hall, Ohmic and Plastic flow depending on the magnetic field and the density of the crust. The horizontal axis is either the radius of the star from the crust-core interface to the neutron drip point or the density (upper x-axis). The vertical axis is the intensity of the magnetic field.}
    \label{FIG:1}
\end{figure}

\subsection{Neutron Star Properties}

For the numerical integration of partial differential equations (\ref{dPsi}) and (\ref{dI}) we require a model for the crust microphysical parameters. We approximate the crust with a spherical shell, starting from the crust-core boundary $R_{CC}$ up to the neutron drip point $R_{ND}$, where $R_{CC}/R_{ND}=0.95$. We use the density, electron number density, elastic limit and conductivity profiles from our previous work, \cite{Lander:2019}. In particular, we define a dimensionless coordinate
\beq
\mathcal{R}\equiv \frac{r-R_{CC}}{R_{ND}-R_{CC}},
\eeq
which is zero at the crust-core boundary and one at the neutron-drip point. We use the analytical approximation for $\rho$ with the scaled radius:
\beq
\tilde\rho\equiv\frac{\rho}{\rho_{CC}}=400\left(1-\frac{R_{CC}}{R_{ND}}\right)^2(1-\mathcal{R})^2+0.004,
\label{ne_eqn}
\eeq
which is close to the exact result in the region between the neutron drip and the crust-core interface, based on the equation of state of  \cite{Douchin:2001}. In the above relation (\ref{ne_eqn}) the ratio of the density at the crust-core interface and the neutron drip point is 250. Based on the same equation of state,  we use analytical fits to  $Z,A,x_{fn}$ and $\rho$ throughout the crust, and calculate the electron number density $n_e$. This result is approximated to good accuracy by the following expression:
\beq
n_e=10^{36}(1.5\tilde\rho^{2/3}+1.9\tilde\rho^2)\textrm{ cm}^{-3}.
\eeq
We assume the electric conductivity scales as $\sigma \propto n_{e}^{2/3}$, and set the conductivity at the base of the crust to $\sigma(R_{CC})=10^{24}$s$^{-1}$, which has the same functional form as the expression used in \cite{GOURGOULIATOS:2014a}. 
The Coulomb parameter is defined as  follows
\beq
\Gamma=\frac{Z^2 e^2}{a_I k_B T}
\eeq
where the ion sphere radius is $a_I=(4\pi n_I/3)^{-1/3}$. For $\Gamma>175$ the envelope of the neutron star starts to crystallise.

For the implementation of our failure criterion, we use the fit of \citet{Chugunov:2010} giving the following expression for the critical stress:
\beq
\tau_{el}=\left(0.0195-\frac{1.27}{\Gamma-71}\right)\frac{Z^2 e^2 n_I}{a_I},
\eeq
which takes the following form once expressed in terms of the scaled density: 
\beq
\tau_{el}=5.1\times 10^{29}(0.4\tilde\rho+0.5\tilde\rho^3)\textrm{ g\ cm}^{-1}\textrm{s}^{-2}.
\eeq
The plastic flow viscosity is a largely unknown quantity. We approximate it as a scaled form of the critical stress:
\beq
\nu=\nu_{0}(0.4\tilde\rho+0.5\tilde\rho^3).
\label{PLASTIC_NU}
\eeq
The scaling parameter is chosen to be 
$\nu_0=2.5\times 10^{38}\textrm{ g\ cm}^{-1}\textrm{s}^{-1}$, using the estimate from \citet{Lander:2016} demanding that the corona of a magnetar is a persistent phenomenon leading to twists lasting for around $10$-yr  \citep{Beloborodov:2007}. We have further explored two more values for the scaling parameter, where the plastic flow viscocity is higher: $\nu_0=2.5\times 10^{39}\textrm{ g\ cm}^{-1}\textrm{s}^{-1}$ and a lower one $2.5\times 10^{37}\textrm{ g\ cm}^{-1}\textrm{s}^{-1}.$

Using equations \eqref{RH} and \eqref{FAILURE_EQ} and the crust profile we have adopted, we can explore the parameter range where the field would be dominated by the Hall evolution, Ohmic decay and the plastic flow; these are plotted in Figure \ref{FIG:1}. In general the deeper part of the crust would be less likely to fail, as $\tau_{el}$ is higher there. Thus the magnetic field evolution in the inner crust is dominated by the Hall effect for fields in the range $2\times 10^{12}-2\times 10^{15}$ G, for fields lower than that the evolution is dominated by the Ohmic decay and for fields higher than that the effect of plastic flow becomes important. Close to the neutron drip point, magnetic field higher than $10^{14}$G will lead to crust failure and a plastic flow, whereas fields below $10^{12}$G are dominated by Ohmic decay in this region. 

The crust density varies from $\sim 10^{14}$g cm$^{-3}$ at the crust-core boundary to $\sim 10^{7}$g cm$^{-3}$ at the surface. The inclusion of regions with such a range of densities would make the calculation extremely slow, as the timestep would be set by the fastest moving electrons, the ones located near the surface of the crust. Furthermore, a realistic treatment of this region requires the inclusion of the interaction between the crust and the magnetosphere \citep{Akgun:2018, Karageorgopoulos:2019}, which is beyond the scope of the current study. For this reason, in these simulations we consider the part of the crust extending from the crust-core interface to the neutron drip point. This part of the crust can also harbour far higher stresses and hence elastic energy than the weak outer crust, and so is likely to be more relevant for powering magnetar outbursts.

Regarding the core, we have adopted two basic approaches. In one of these we regard the magnetic field as having been expelled from the core and confined in the crust. In the other, some poloidal flux penetrates into the core, but the core field does not evolve within the timeframe of the simulation. Similarly to the crust-magnetospheric interplay, the physics of the crust-core interaction involve phenomena such as ambipolar diffusion and the evolution of the magnetic field within a superfluid and superconducting that deserve separate detailed studies \citep{Lander:2013a, Lander:2013b, Passamonti:2017a, Passamonti:2017b}. Therefore, in the current study we assume that the field of the core is a fixed boundary condition.

\section{Simulations}
\label{SIMULATIONS}

\begin{table*}
\centering
\caption{Numerical models implemented for a magnetic field that is confined in the crust. The first column is the name of the run, subsequent columns are the value of $\Psi_0$, the value of the field at the neutron star pole, the energy in the crust, whether the field threads the core or not, the value of $\nu_0$ and the type of failure  of the run.}
\label{TAB:1}
\begin{tabular}{ccccccc}
\hline
Name	& $\Psi_0$ 	&$B_{dip,0}$ ($10^{14}$ G)& $E_{mag}~(10^{46}{\rm erg})$ & Core &$\nu_0$(g cm$^{-1}{\rm  s^{-1}}$) & Failure  \\
\hline
LC-1 & $200$& $1$& $3.1$ & No & $2.5\times 10^{39}$ & Local \\
LC-2 &$200$& $1$& $3.1$ & No & $2.5\times 10^{38}$ & Local \\
LC-3 &$200$& $1$& $3.1$ & No & $2.5\times 10^{37}$ & Local \\
\hline
LT-1 & $5$& $0.5$& $1.95$ & Yes & $2.5\times 10^{39}$ & Local \\
LT-2 &$5$& $0.5$& $1.95$ & Yes & $2.5\times 10^{38}$ & Local \\
LT-3 & $10$& $1$& $7.8$ & Yes & $2.5\times 10^{39}$ & Local \\
LT-4 & $10$& $1 $& $7.8$ & Yes & $2.5\times 10^{38}$ & Local \\
\hline
IC-1 & $200$& $1$& $3.1$ & No & $2.5\times 10^{39}$ & Intermediate \\
IC-2 &$200$& $1$& $3.1$ & No & $2.5\times 10^{38}$ & Intermediate \\
IC-3 &$200$& $1$& $3.1$ & No & $2.5\times 10^{37}$ & Intermediate \\
\hline
IT-1 & $5$& $0.5$& $1.95$ & Yes & $2.5\times 10^{39}$ & Intermediate \\
IT-2 &$5$& $0.5$& $1.95$ & Yes & $2.5\times 10^{38}$ & Intermediate \\
IT-3 & $10$& $1$& $7.8$ & Yes & $2.5\times 10^{39}$ & Intermediate  \\
IT-4 & $10$& $1$& $7.8$ & Yes & $2.5\times 10^{38}$ & Intermediate \\
\hline
GC-1 & $200$& $1$& $3.1$ & No & $2.5\times 10^{39}$ & Global \\
GC-2 &$200$& $1$& $3.1$ & No & $2.5\times 10^{38}$ & Global \\
GC-3 &$200$& $1$& $3.1$ & No & $2.5\times 10^{37}$ & Global \\
\hline
GT-1 & $5$& $0.5$& $1.95$ & Yes & $2.5\times 10^{39}$ & Global \\
GT-2 &$5$& $0.5$& $1.95$ & Yes & $2.5\times 10^{38}$ & Global \\
GT-3 & $10$& $1$& $7.8$ & Yes & $2.5\times 10^{39}$ & Global \\
GT-4 & $10$& $1$& $7.8$ & Yes & $2.5\times 10^{38}$ & Global \\
\hline
HC-1 & $200$& $1$& $3.1$ & No & N/A & No  \\
\hline
HT-1 & $5$& $0.5$& $1.95$ & Yes & N/A & No  \\
HT-2 & $10$& $1 $& $7.8$ & Yes & N/A & No \\
\hline
\end{tabular}
\end{table*}

\subsection{Numerical Setup}

We have discretised the numerical domain in radius in $r$ and $\mu=\cos\theta$. The typical resolution we use is $100^2$, but we have also experimented with higher resolution runs ($200^2$) to confirm the validity of our results and numerical convergence. 

The simulation consists of a main loop for the time integration of the partial differential equations (\ref{dPsi}) and (\ref{dI}). Within the main loop, we test whether the failure criterion, Equation \eqref{FAILURE_EQ}, is satisfied or not. Should the failure criterion be satisfied, then depending on whether we consider a local, intermediate or global failure, we integrate (\ref{Laplacian}) in the appropriate domain and apply the relevant boundary conditions. We evaluate the plastic flow velocity, using the Gauss-Seidel iterative method, until it relaxes to a solution. This integration is the main bottleneck of the calculation, as it requires a large number of steps for the convergence of equation (\ref{Laplacian}). In practice, of the order $10^{4}$ iterations are needed for this calculation for a $100^2$ resolution, leading to a drastic increase of the integration time by four orders of magnitude if we update the plastic flow velocity at every single time-step. Fortunately we have found that this is not necessary, and updating every 10-100 steps is typically sufficient, since the plastic flow velocity -- being sourced by changes in the magnetic field -- changes only slowly with time. Simulations are thus typically a factor of 100-1000 slower than the same model under Hall-Ohmic evolution. Once the plastic flow velocity is known, we use the Adams-Bashforth 2nd-order method to integrate in time. Furthermore, we use an adaptable time-step evaluated by a Courant condition \citep{Courant:1952}, which is based on the maximum velocity of the system accounting both for the electron fluid and the plastic flow velocity. 

Spatial derivatives are evaluated using a central difference scheme, a three-point stencil for the second derivative, and five-point stencil for the third derivative. The boundary conditions implemented are those of a multipolar current-free poloidal magnetic field on the surface of the star $(r=R_{ND})$ and with the poloidal current set to zero $I(R_{ND}, \theta)=0$, enforcing that no current flows from the star to the magnetosphere. On the axis of the star we set $\Psi(r, 0)=\Psi(r, \pi)=I(r,0)=I(r,\pi)=0$. The inner boundary condition for the poloidal current is $\frac{d I}{dr}=0$. The poloidal flux boundary condition is either $\Psi(R_{CC}, \theta)=0$ for the regime where the magnetic field has been expelled from the core or fixed to a given function at the initial conditions $\Psi(R_{CC}, \theta)=f(\theta)$, and not allowed to evolve in the core and at the crust-core boundary. 

Regarding the plastic flow, the boundary conditions employed depend on the type of failure we have assumed. In the local failure simulations, the flow velocity is set to zero at the points where the failure criterion is not satisfied. In the intermediate and global flow case, we impose the following boundary conditions for the plastic flow velocity in the solution of equation (\ref{Laplacian}). We set $v_{pl}(r, 0)=v_{pl}(0, \pi)=0$ on the axis and at the crust-core interface $v_{pl}(R_{CC}, \theta)=0$; and at the surface of the integration domain we set free-slip boundary conditions $\partial_r v_{pl}(R_{ND}, \theta)=0$.

\subsection{Initial Conditions}

We have adopted the following initial conditions depending on whether the field threads the core or is confined in the crust. The fields that thread into the core have the following profile for $\Psi$:
\begin{eqnarray}
\Psi_{1} =\Psi_{0}\frac{r}{R_{ND}}\left(1.05-\frac{r}{R_{ND}}\right)\sin^2\theta\, ,
\end{eqnarray}
whereas the ones that are confined in the crust have:
\begin{eqnarray}
\Psi_{2} =\Psi_{0} \left(1.05-\frac{r}{R_{ND}}\right)\left(\frac{r}{R_{ND}}-\frac{R_{CC}}{R_{ND}}\right) \sin^2\theta\,.
\end{eqnarray}
We set the intensity of the magnetic field by the $\Psi_{0}$ parameter. 

For all the families of plastic flow simulations we have performed, we have also simulated a case where only evolution through the Hall-Ohmic effect is allowed, for comparison.

\section{Results}

\label{RESULTS}

%%%%%%%%%%%%%%%%%%%%%%%%%%%%%%%%%%%%%%%%%%
\begin{figure*}
    a\includegraphics[width=0.3\textwidth]{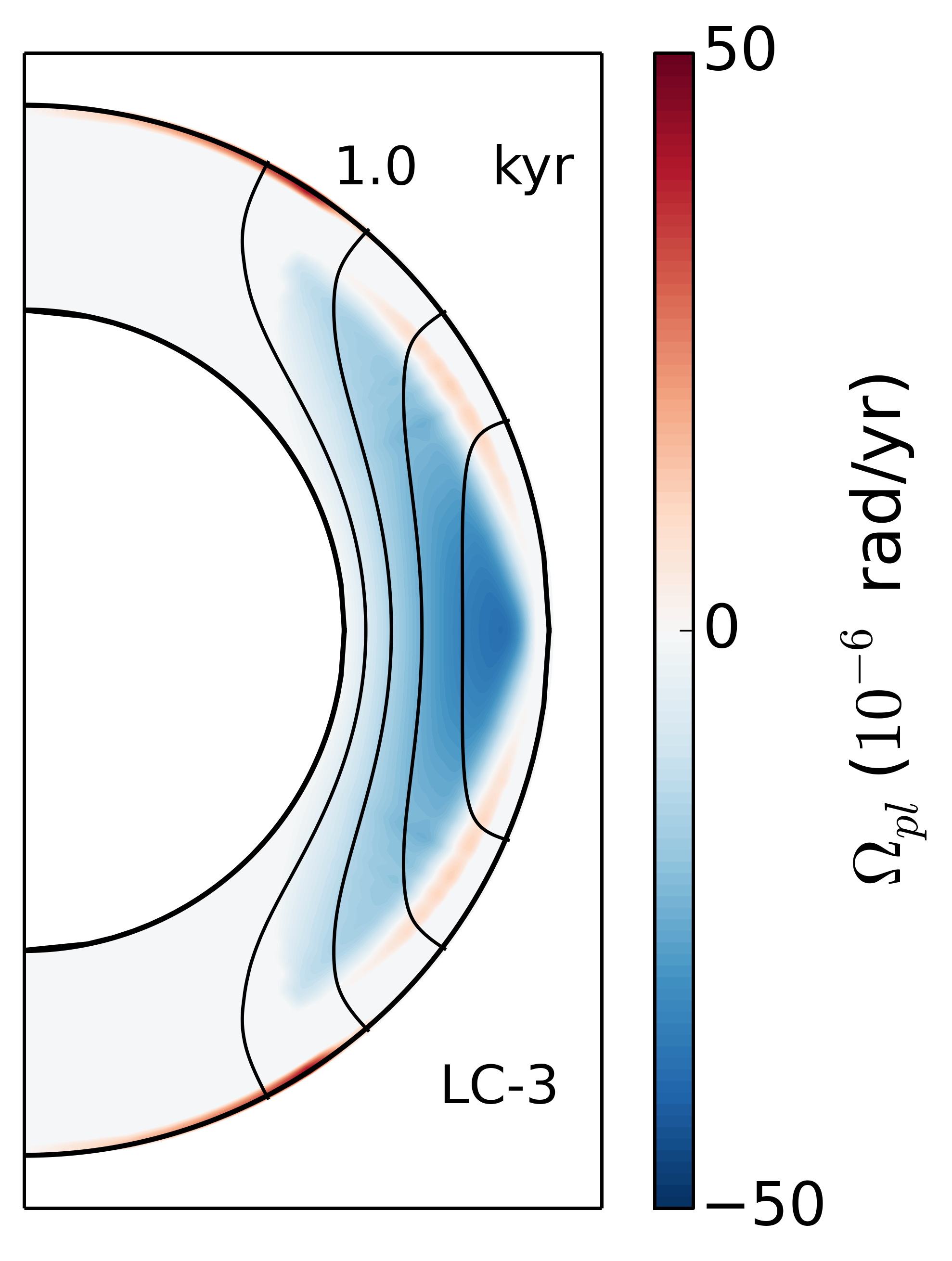}
    b\includegraphics[width=0.3\textwidth]{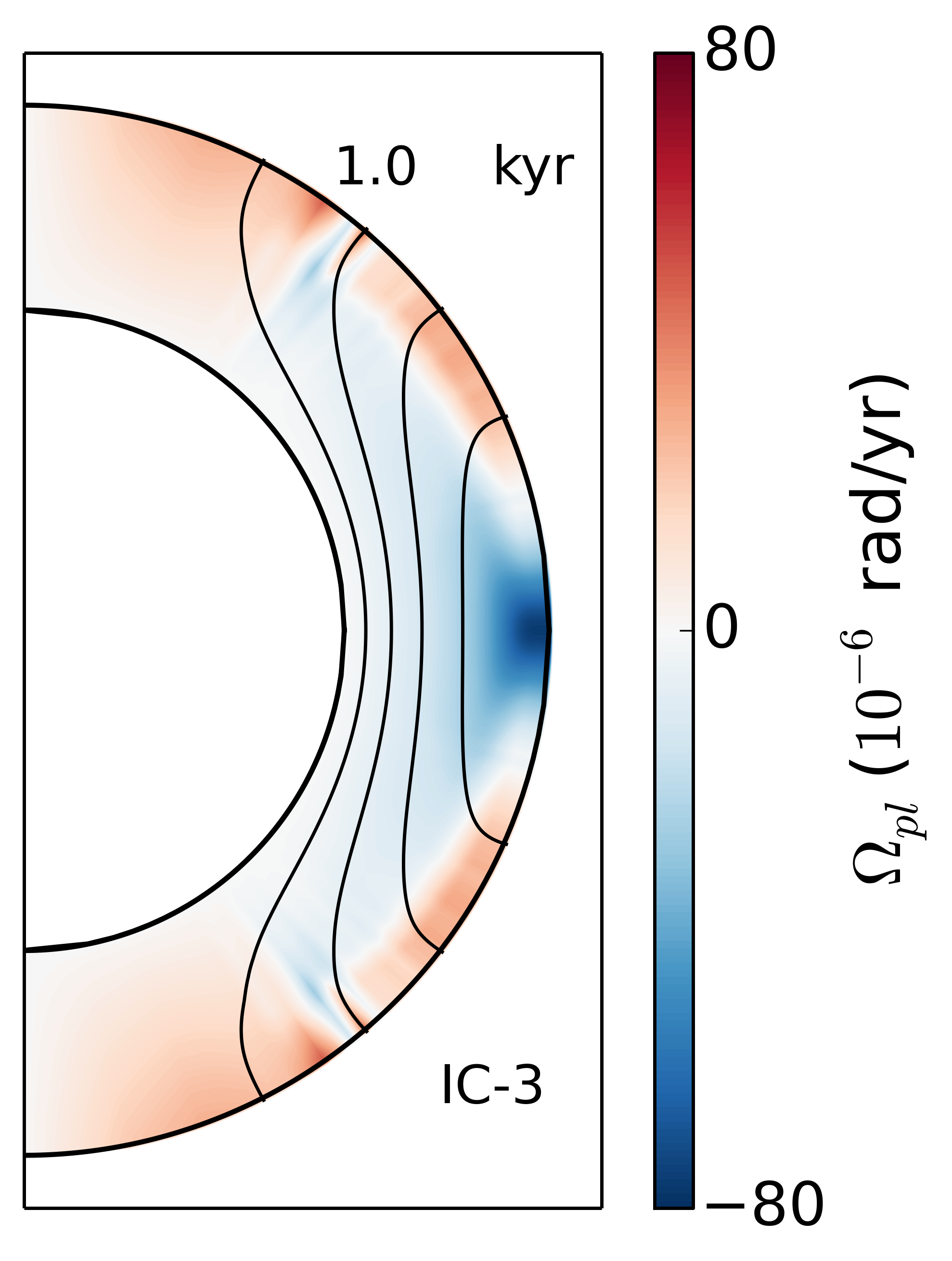}
    c\includegraphics[width=0.315\textwidth]{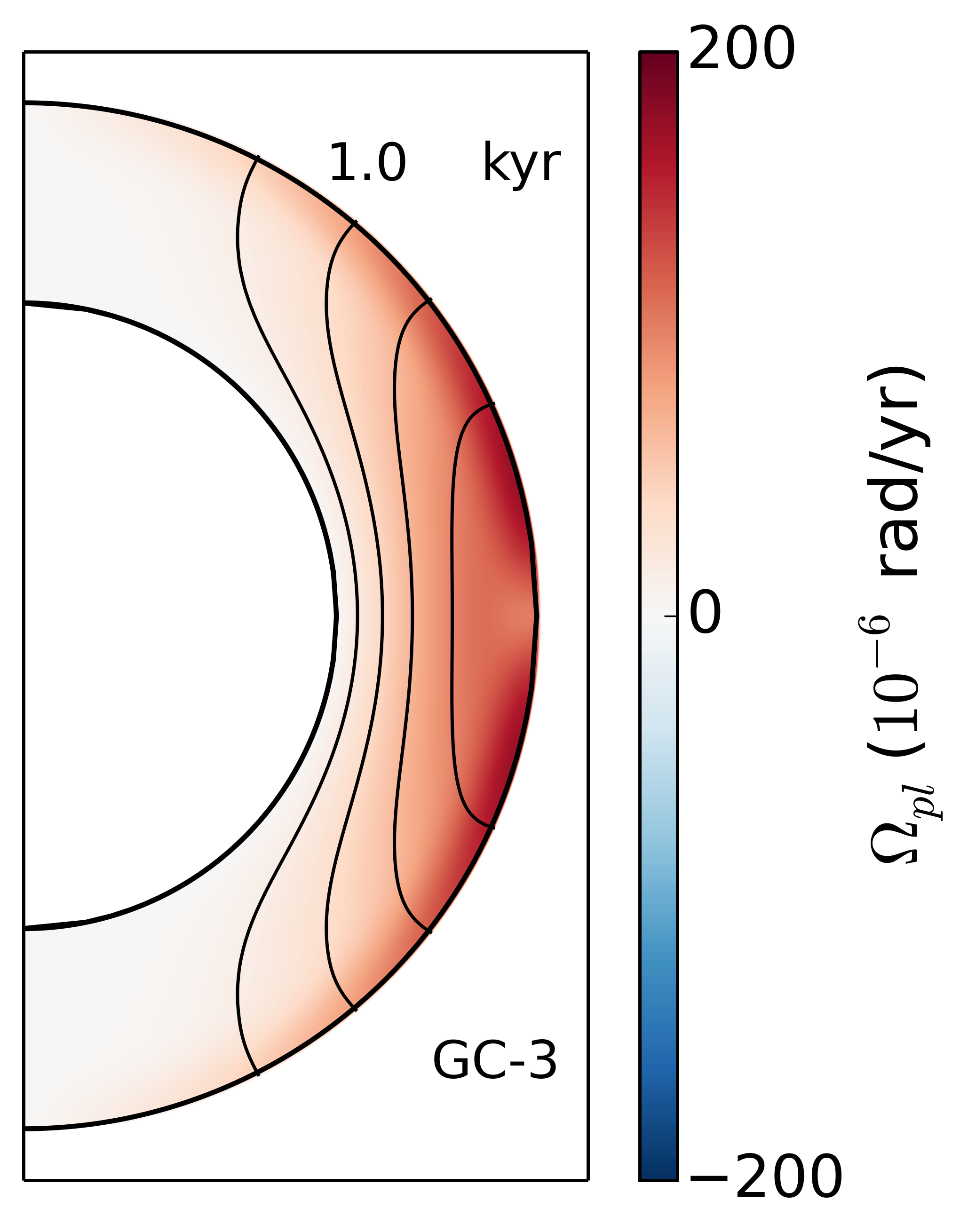}
    \caption{Plots of the poloidal magnetic field line structure (shown in black) and the plastic flow angular velocity (shown in colour), for models where crustal failure is local (a), intermediate (b), or global (c), all after 1 kyr of evolution. The specific runs used are LC-3 (a), IC-3 (b) and GC-3 (c).}
    \label{FIG:2}
\end{figure*}
%%%%%%%%%%%%%%%%%%%%%%%%%%%%%%%%%%%%%%%%%%
\begin{figure}
    a\includegraphics[width=0.24\textwidth]{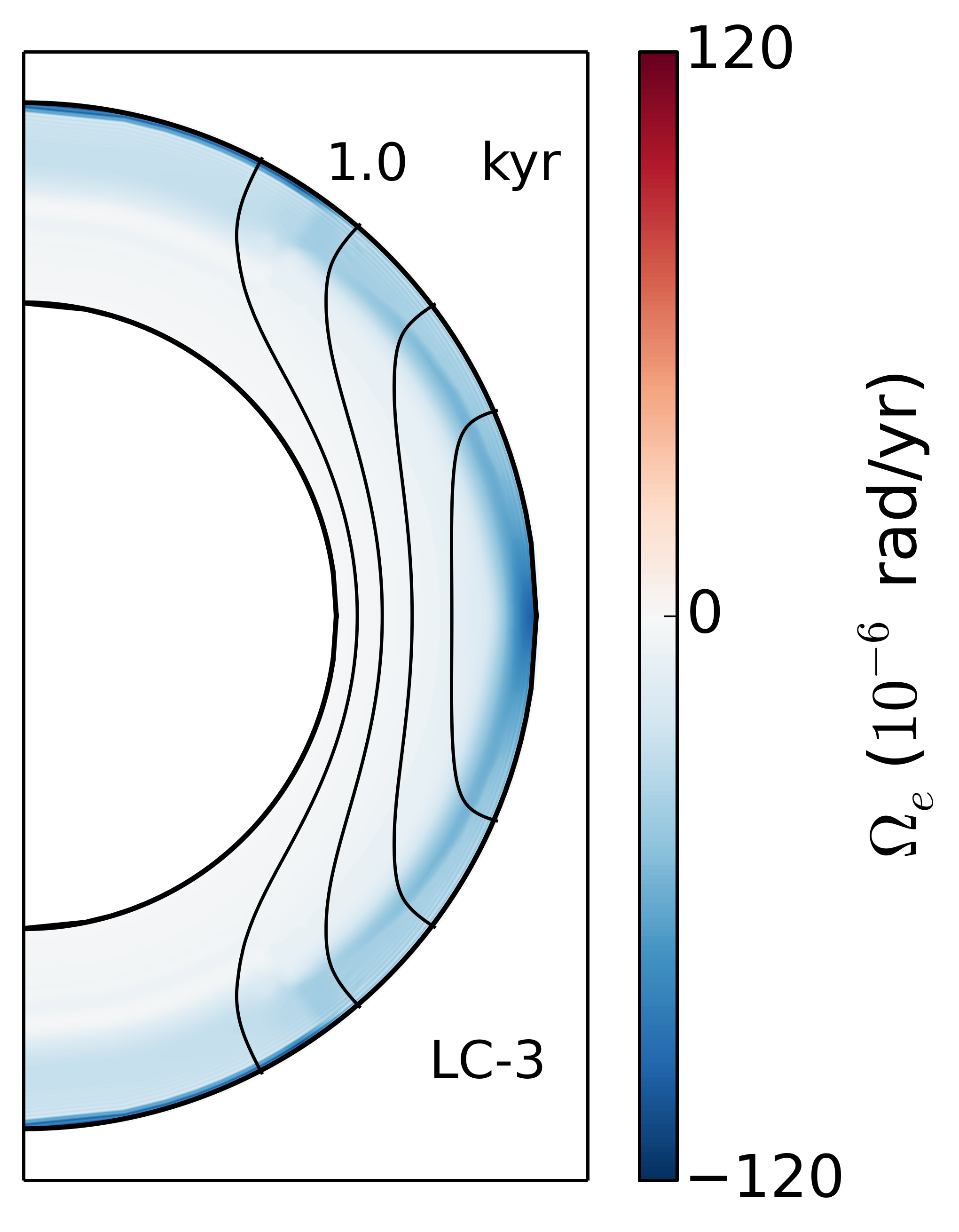}
    b\includegraphics[width=0.205\textwidth]{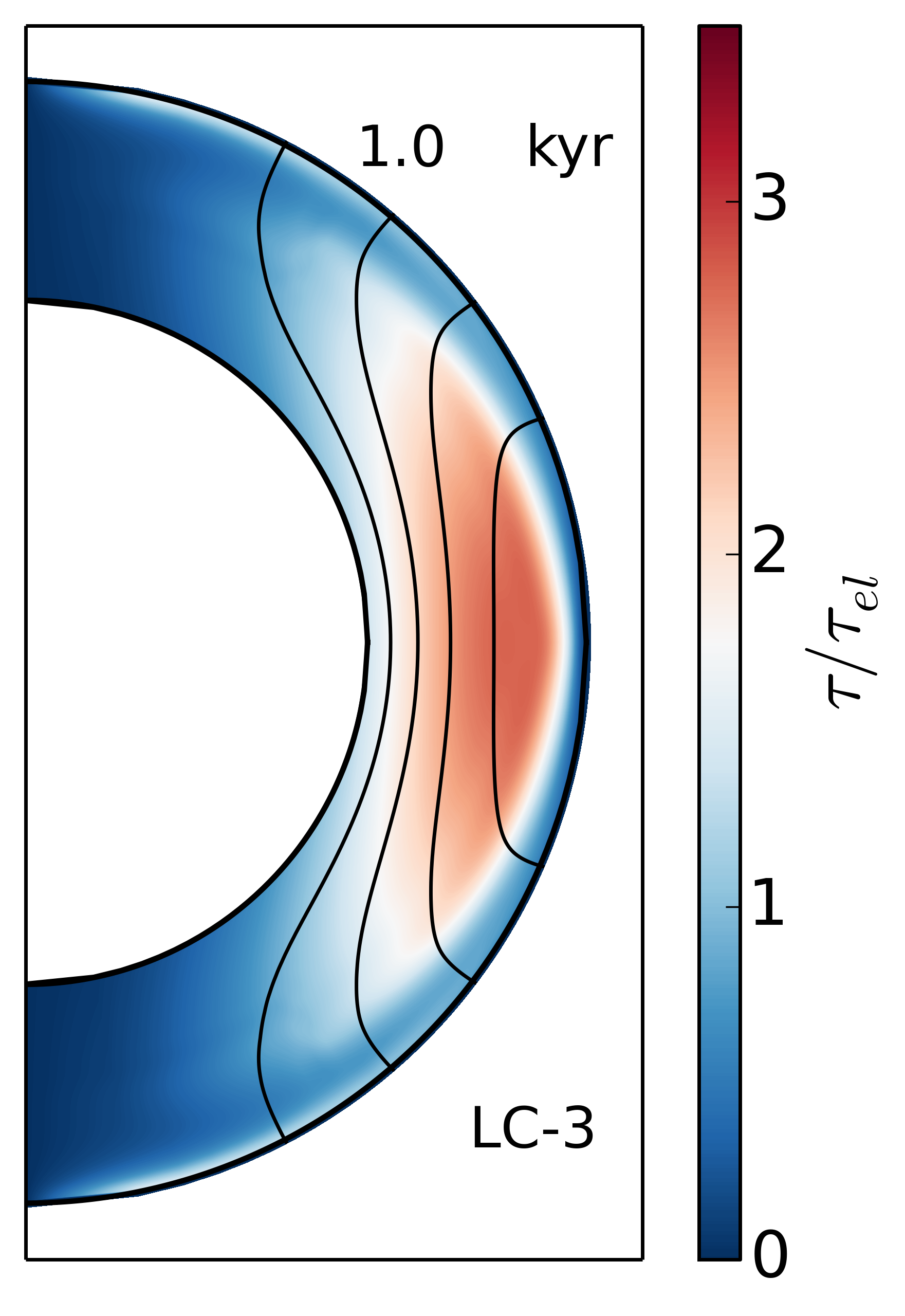}
    \caption{Plots of the electron fluid angular velocity (a) and the ratio of the stress to the critical value (b), in colour, with the poloidal magnetic field lines shown in black, for model LC-3, at 1 kyr.}
    \label{FIG:3}
\end{figure}
%%%%%%%%%%%%%%%%%%%%%%%%%%%%%%%%%%%%%%%%%%
\begin{figure*}
    a\includegraphics[width=0.23\textwidth]{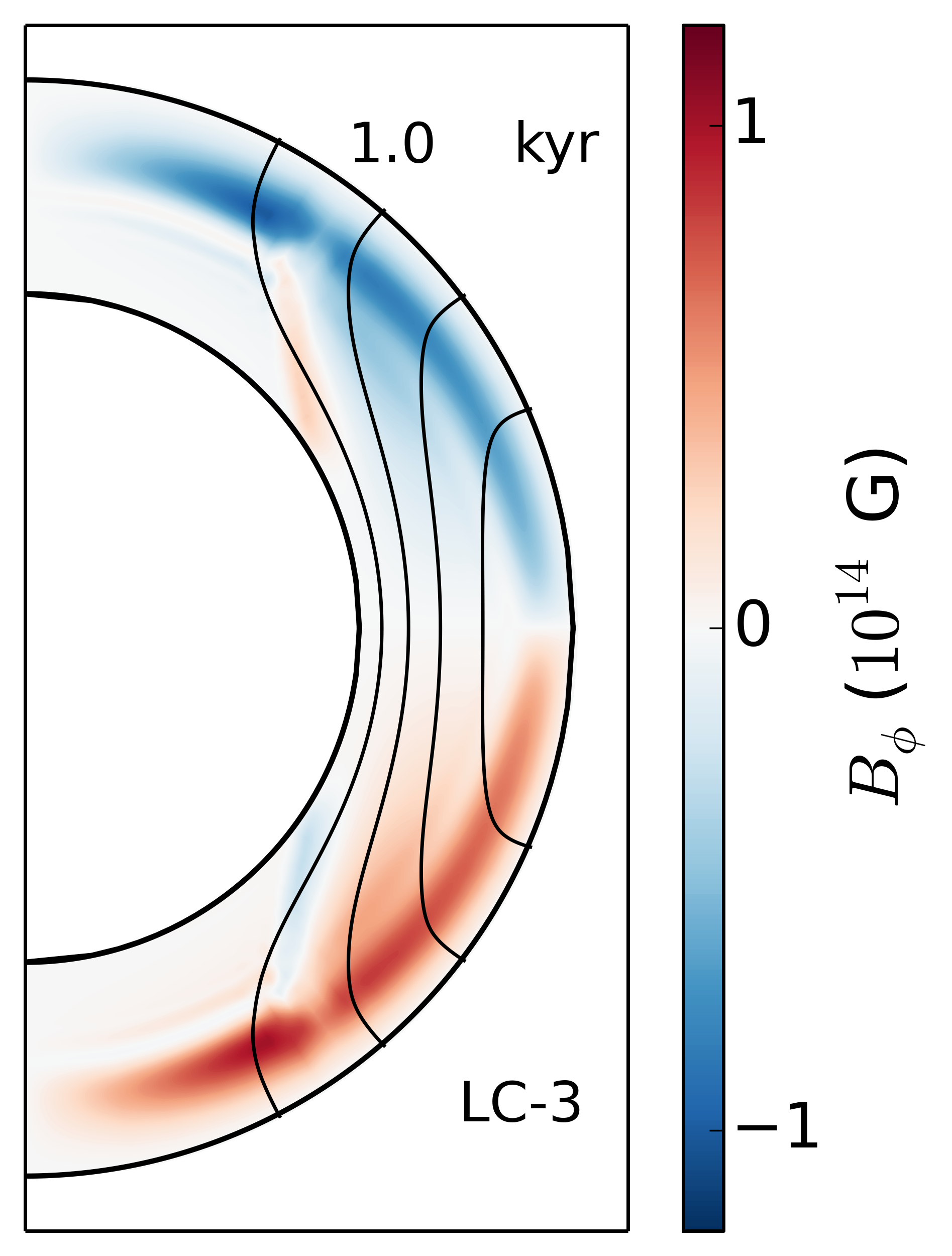}
    b\includegraphics[width=0.23\textwidth]{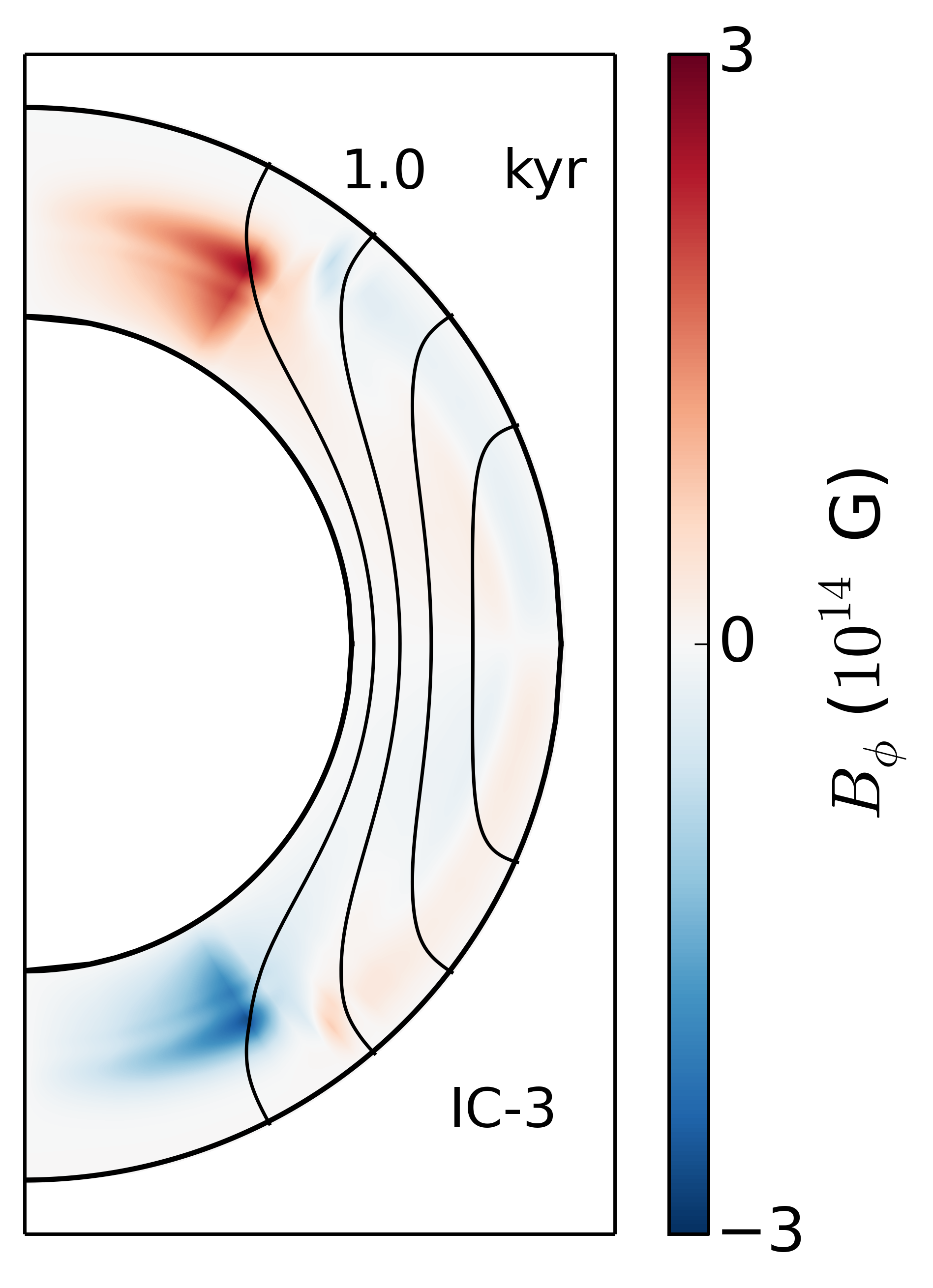}
    c\includegraphics[width=0.245\textwidth]{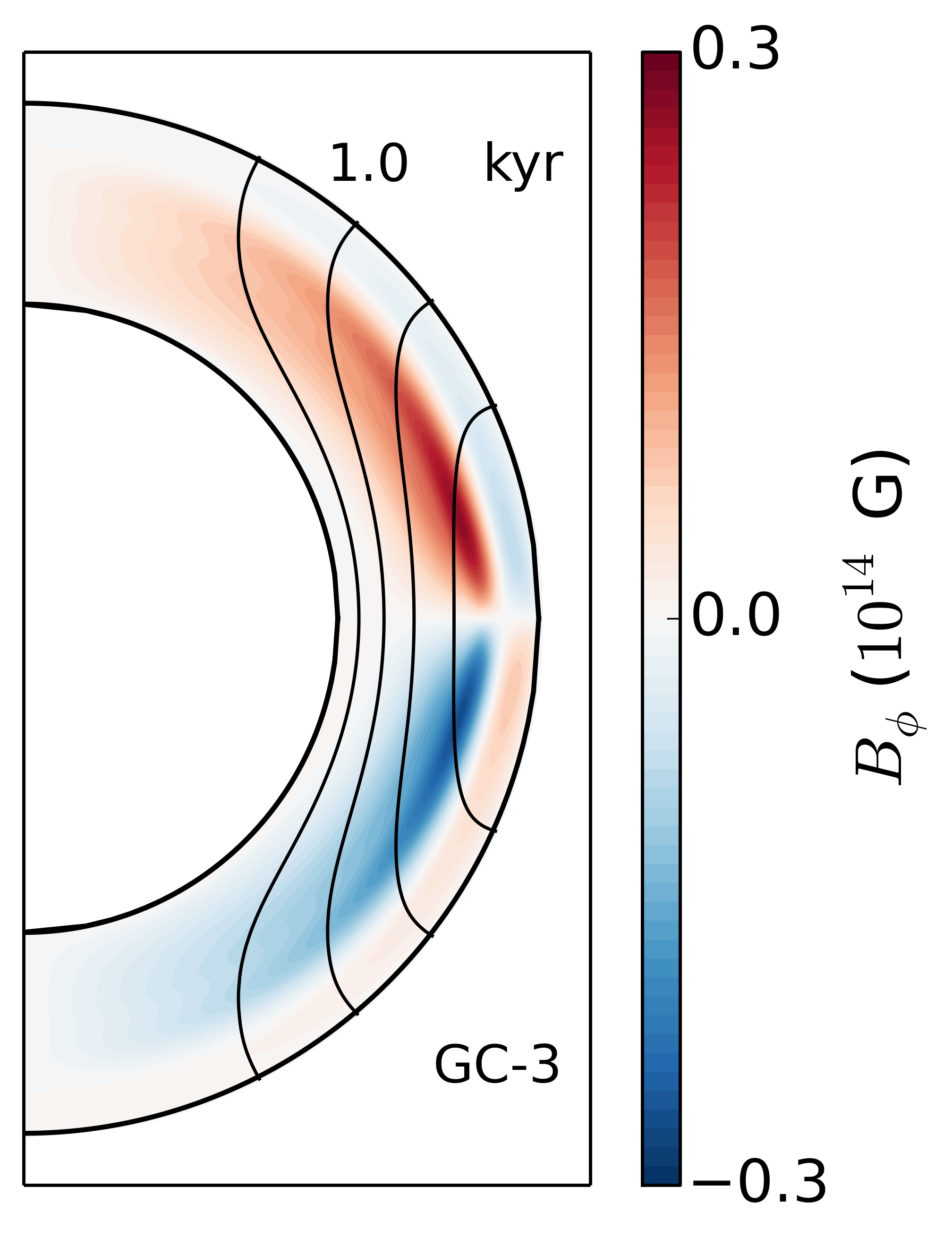}
    d\includegraphics[width=0.23\textwidth]{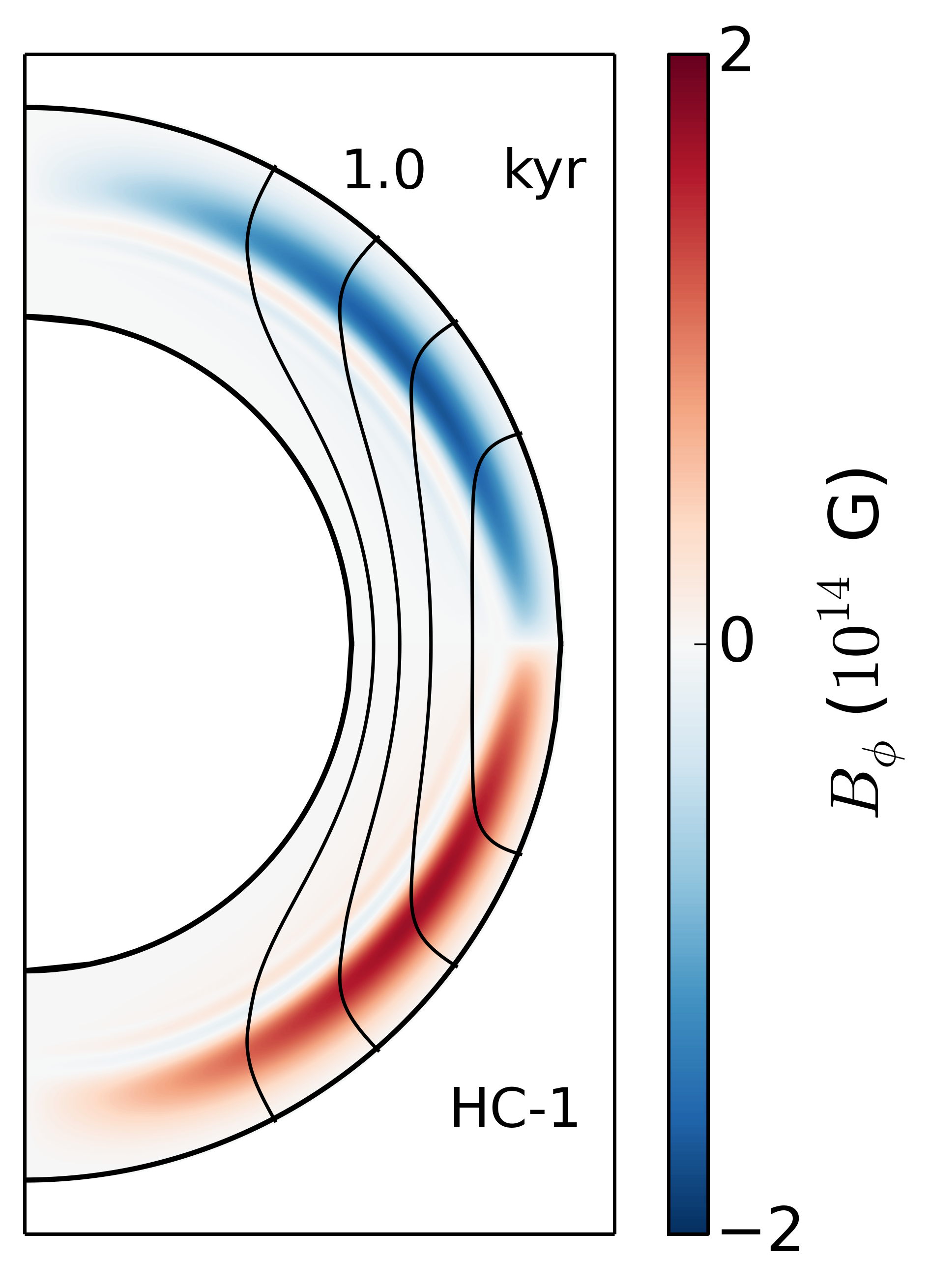}
    e\includegraphics[width=0.23\textwidth]{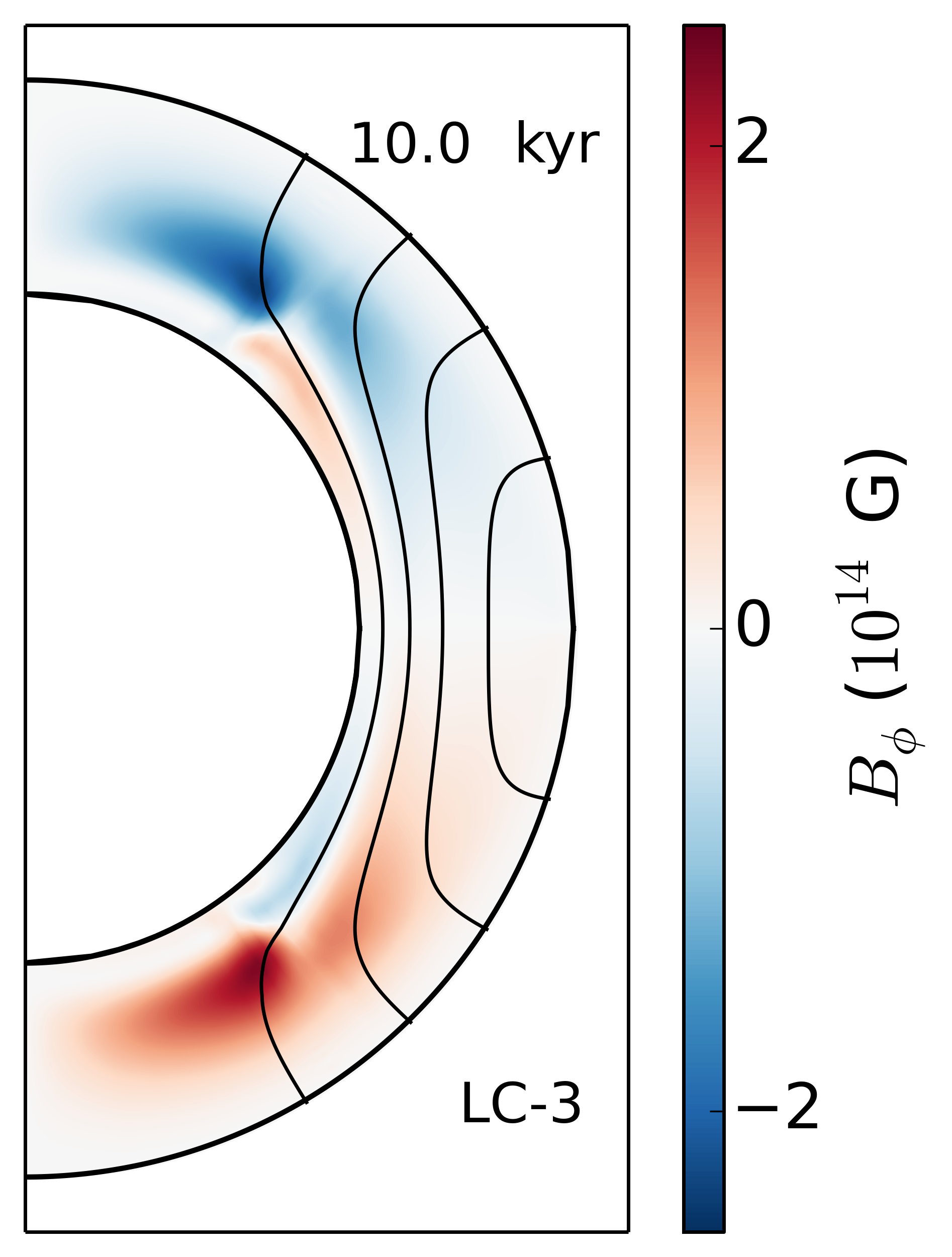}
    f\includegraphics[width=0.23\textwidth]{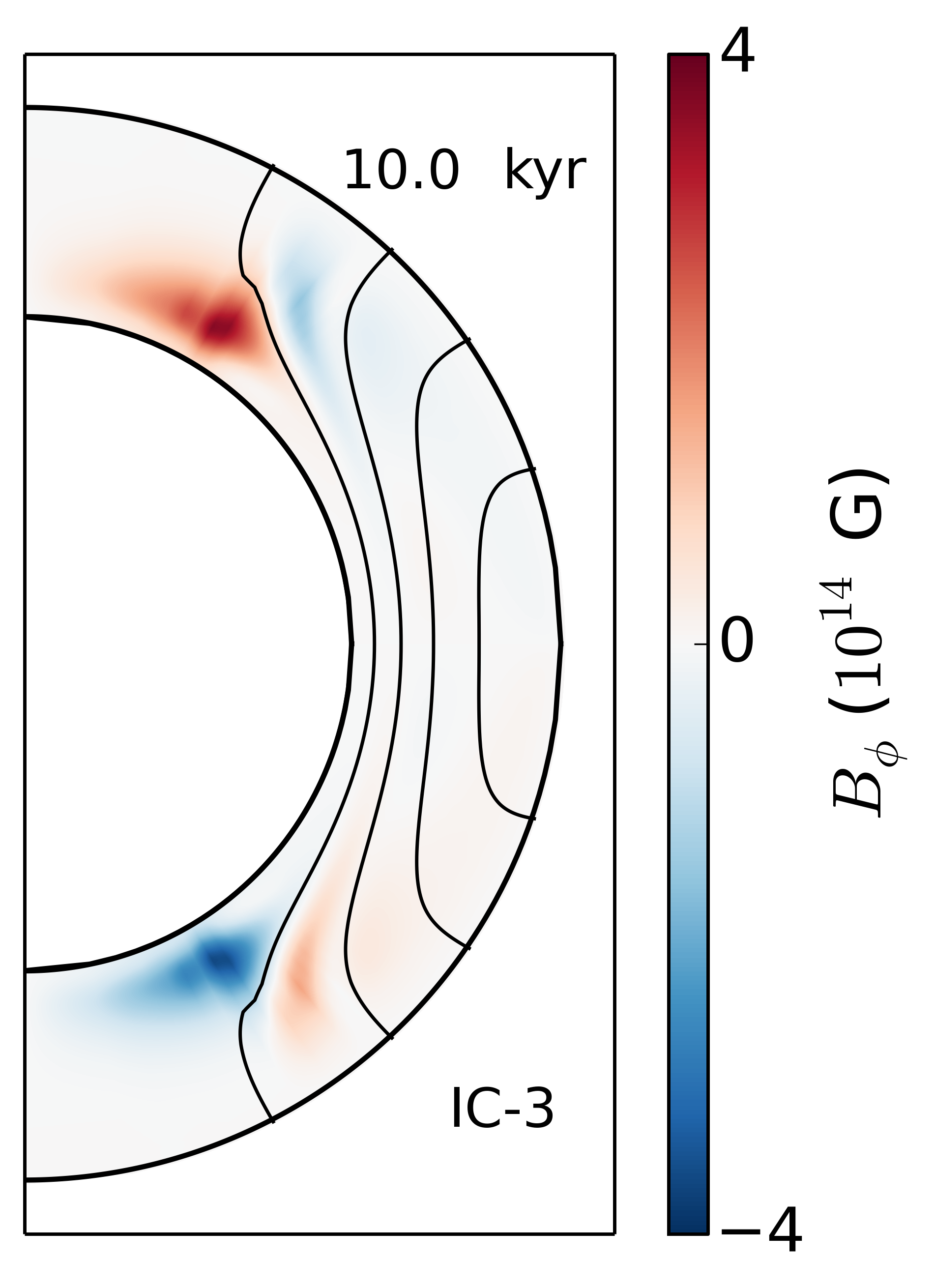}
    g\includegraphics[width=0.245\textwidth]{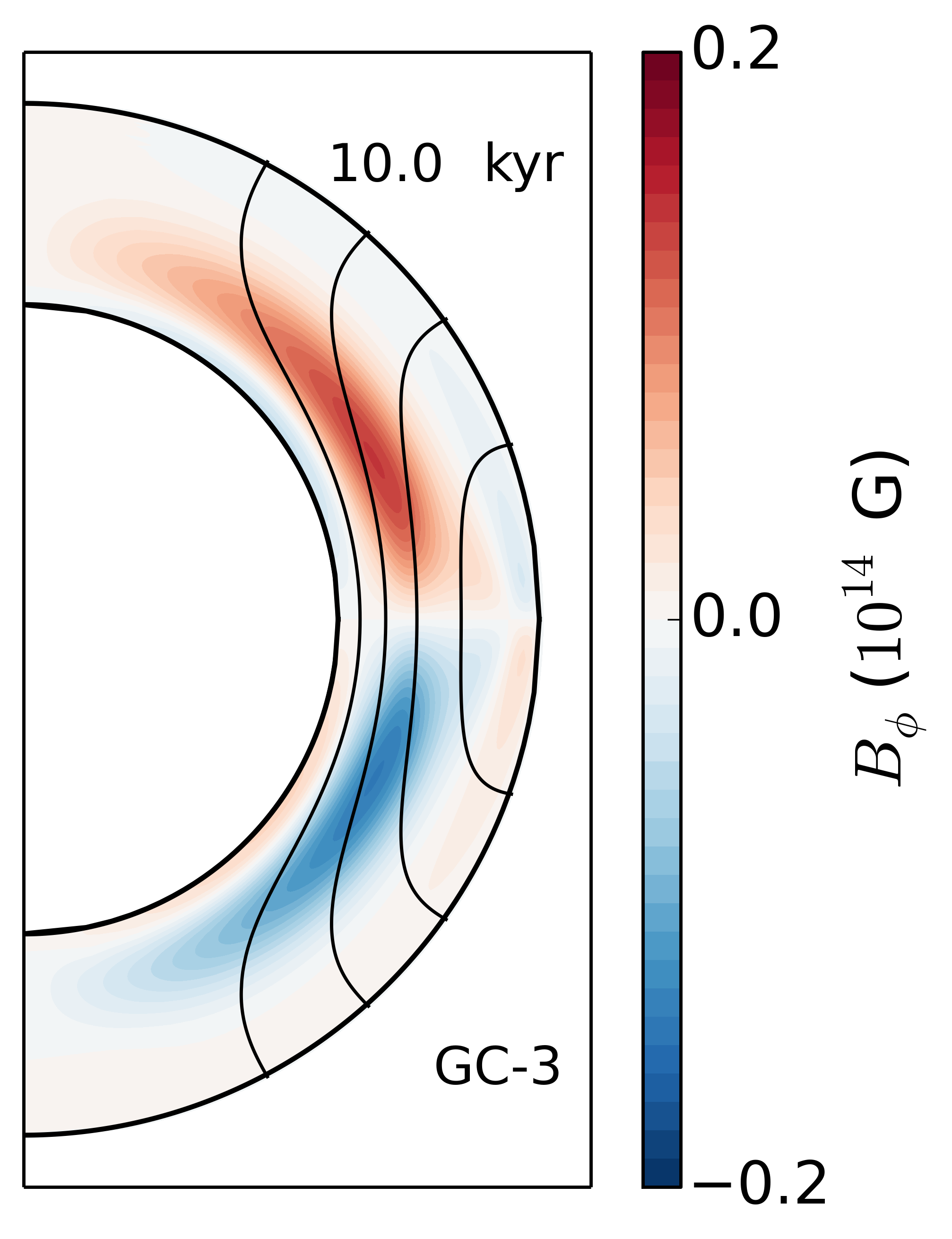}
    h\includegraphics[width=0.23\textwidth]{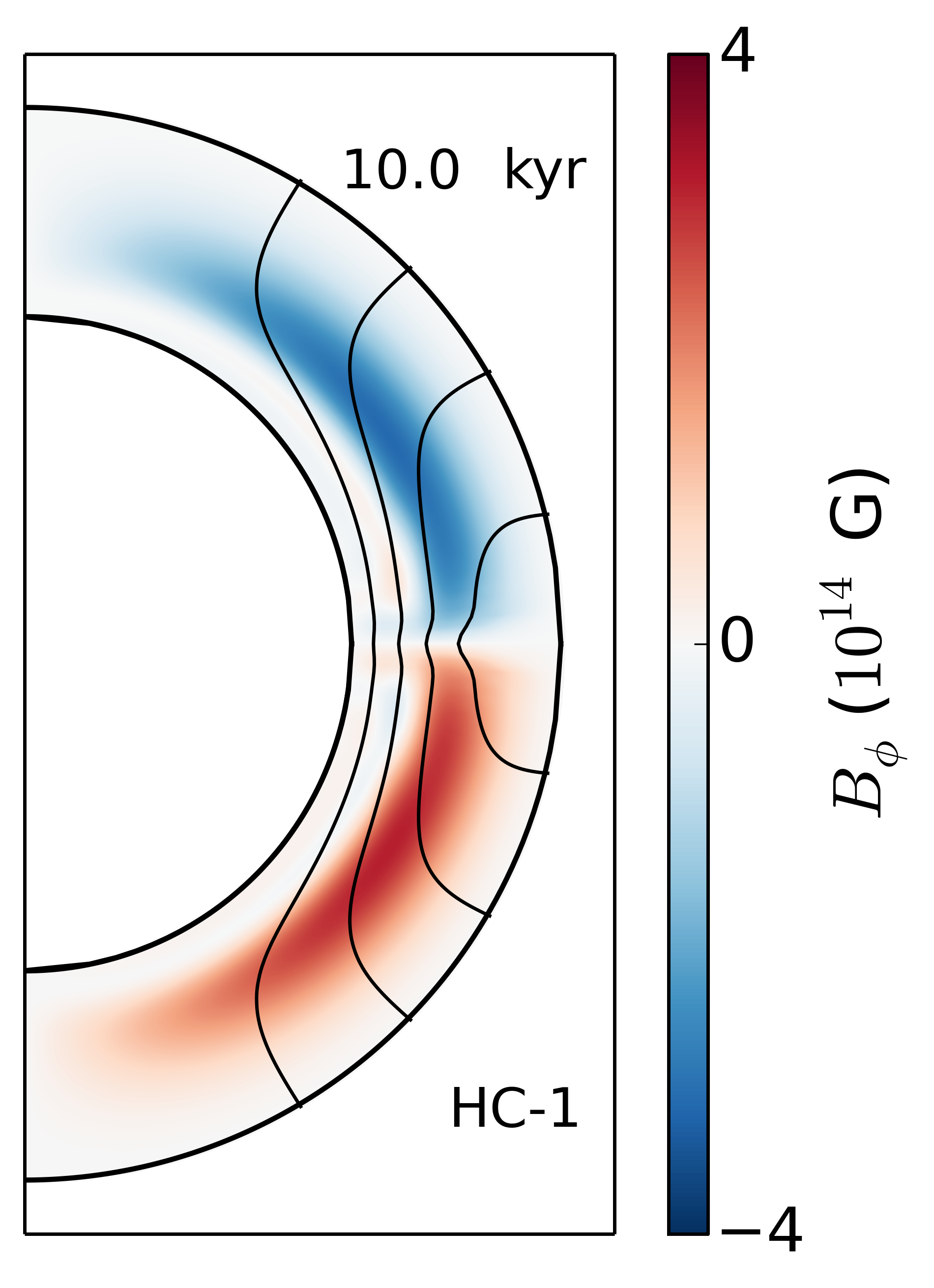}
    \caption{Plots of the poloidal magnetic field lines shown in black and the toroidal field in colour, at 1 kyr (top row) and 10 kyrs (bottom row) for models LC-3 (a and e), IC-3 (b and f), GC-3 (c and g) and HC-1 (d and h). }
    \label{FIG:4}
\end{figure*}
%%%%%%%%%%%%%%%%%%%%%%%%%%%%%%%%%%%%%%%%%%
\begin{figure*}
    a\includegraphics[width=0.3\textwidth]{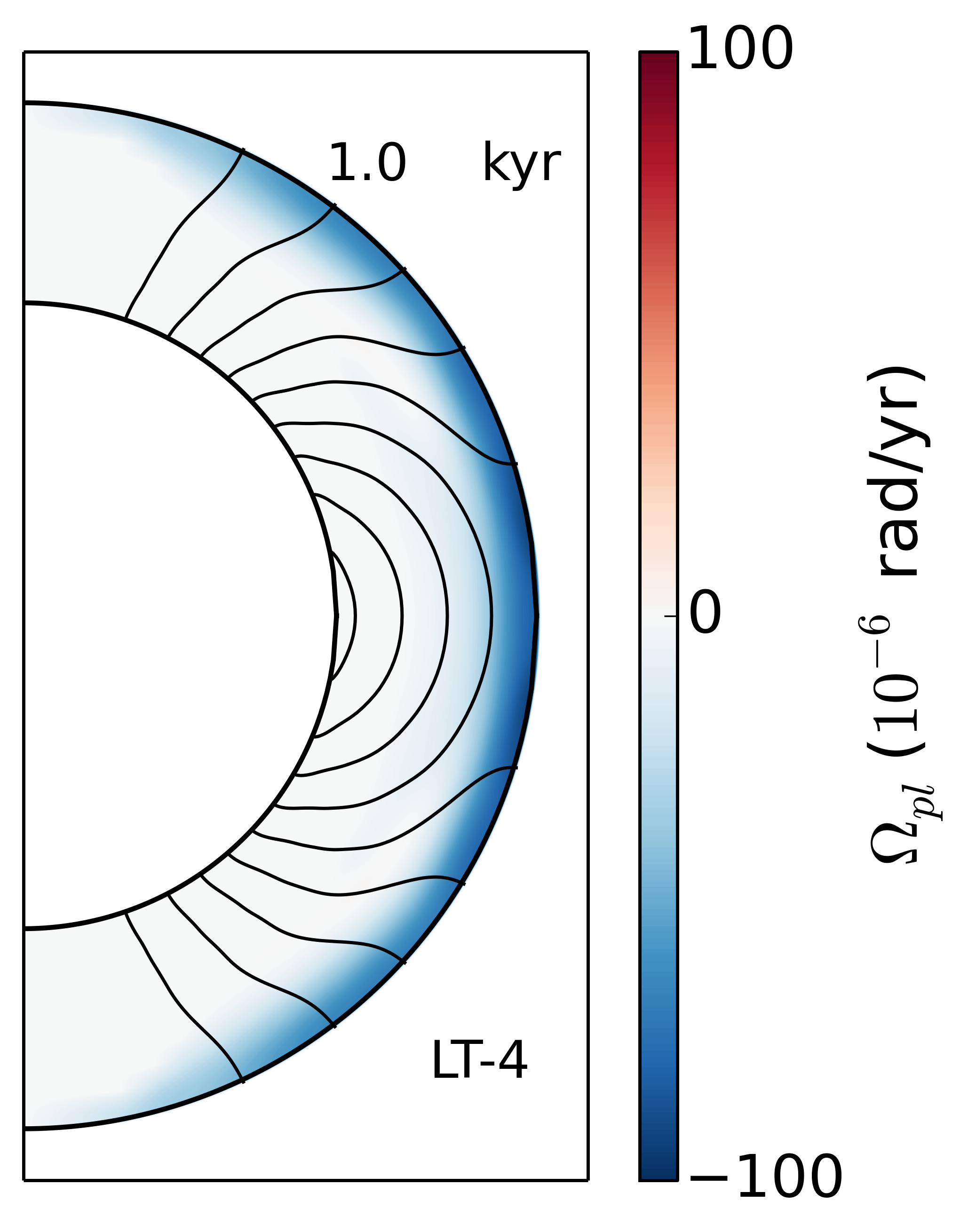}
    b\includegraphics[width=0.3\textwidth]{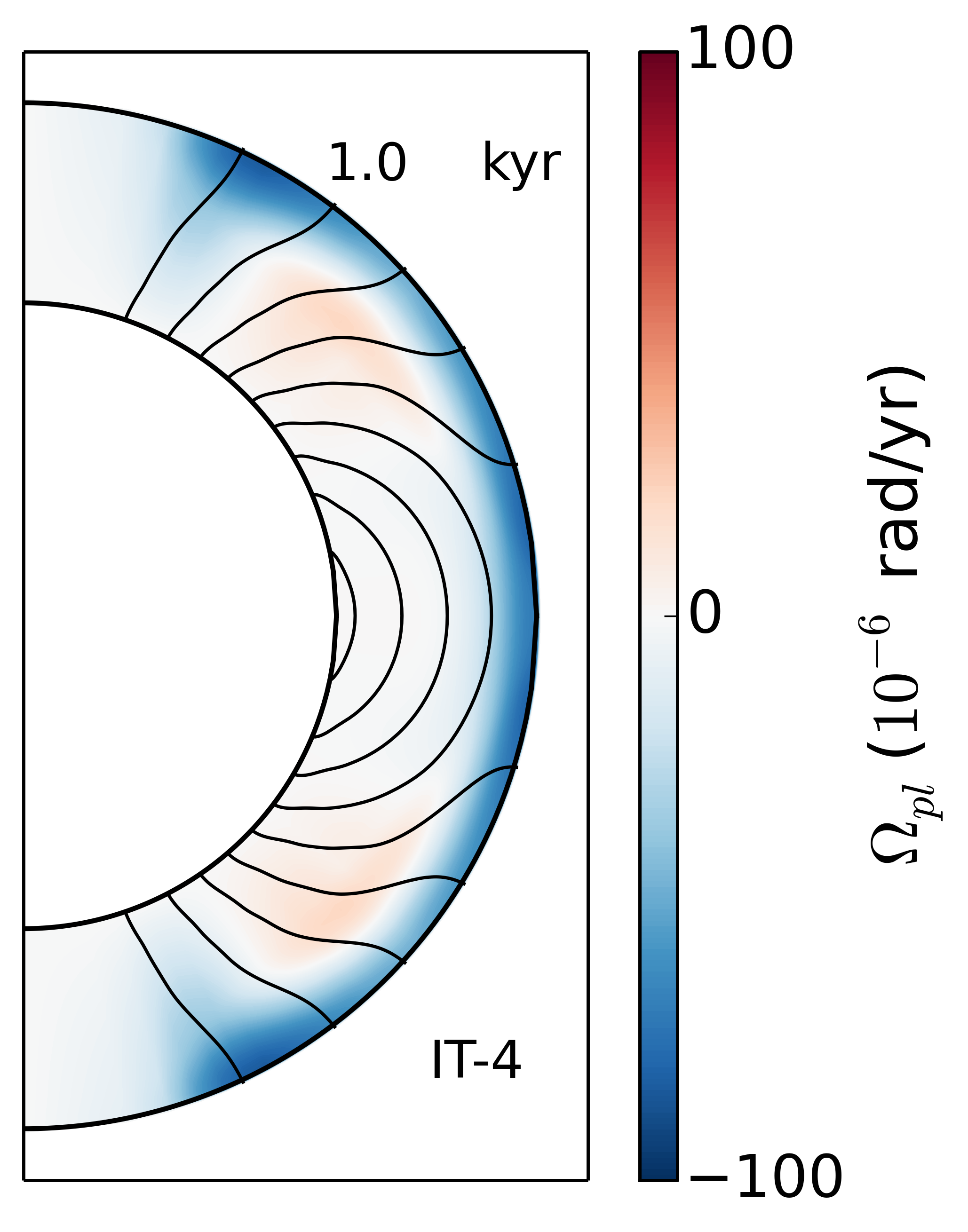}
    c\includegraphics[width=0.3\textwidth]{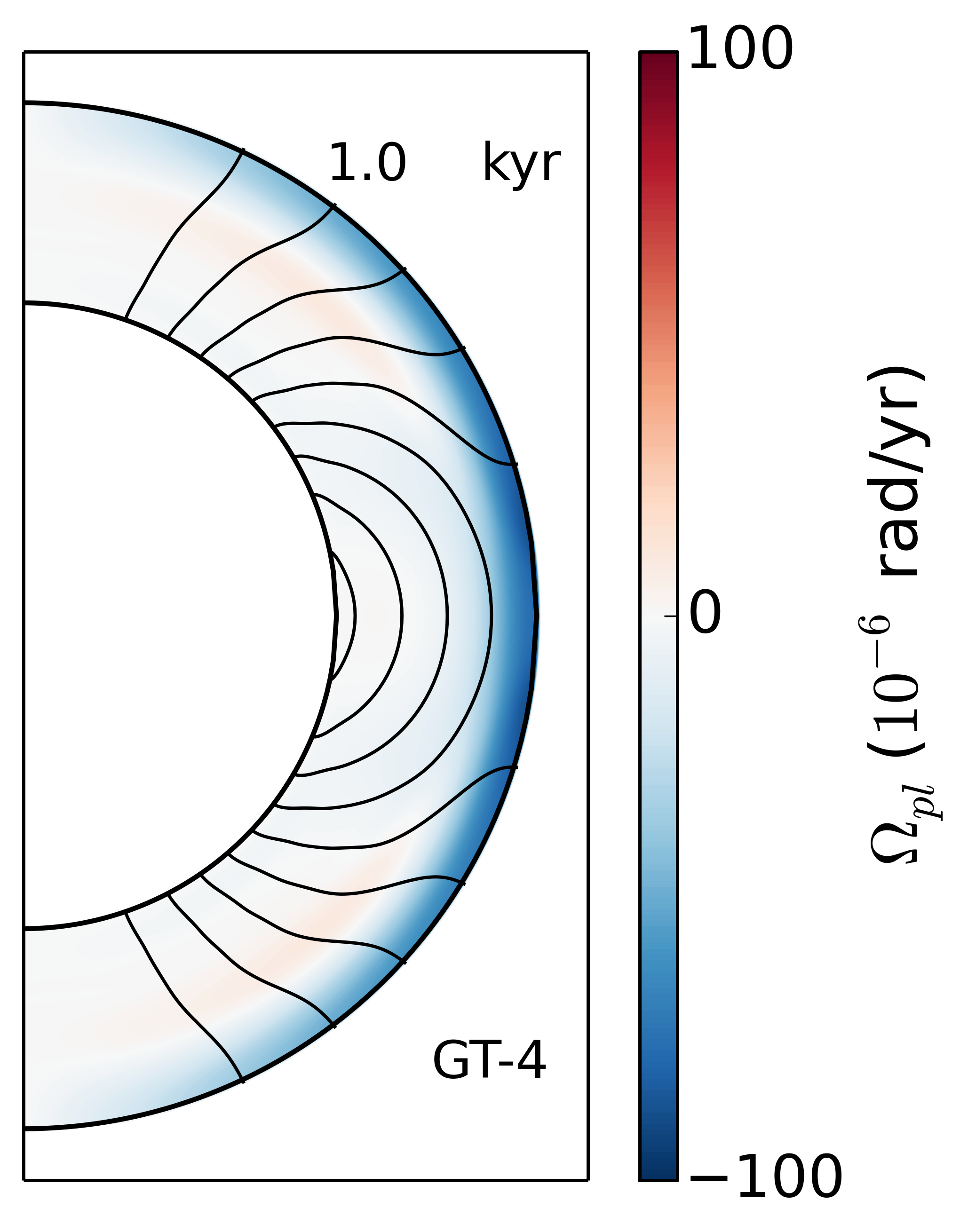}
    \caption{Plots of the poloidal magnetic field lines structure shown in black and the plastic flow angular velocity shown in colour, for models LT-4 (a), IT-4 (b) and GT-4 (c), at 1 kyr. }
    \label{FIG:5}
\end{figure*}
%%%%%%%%%%%%%%%%%%%%%%%%%%%%
\begin{figure}
    \includegraphics[width=0.24\textwidth]{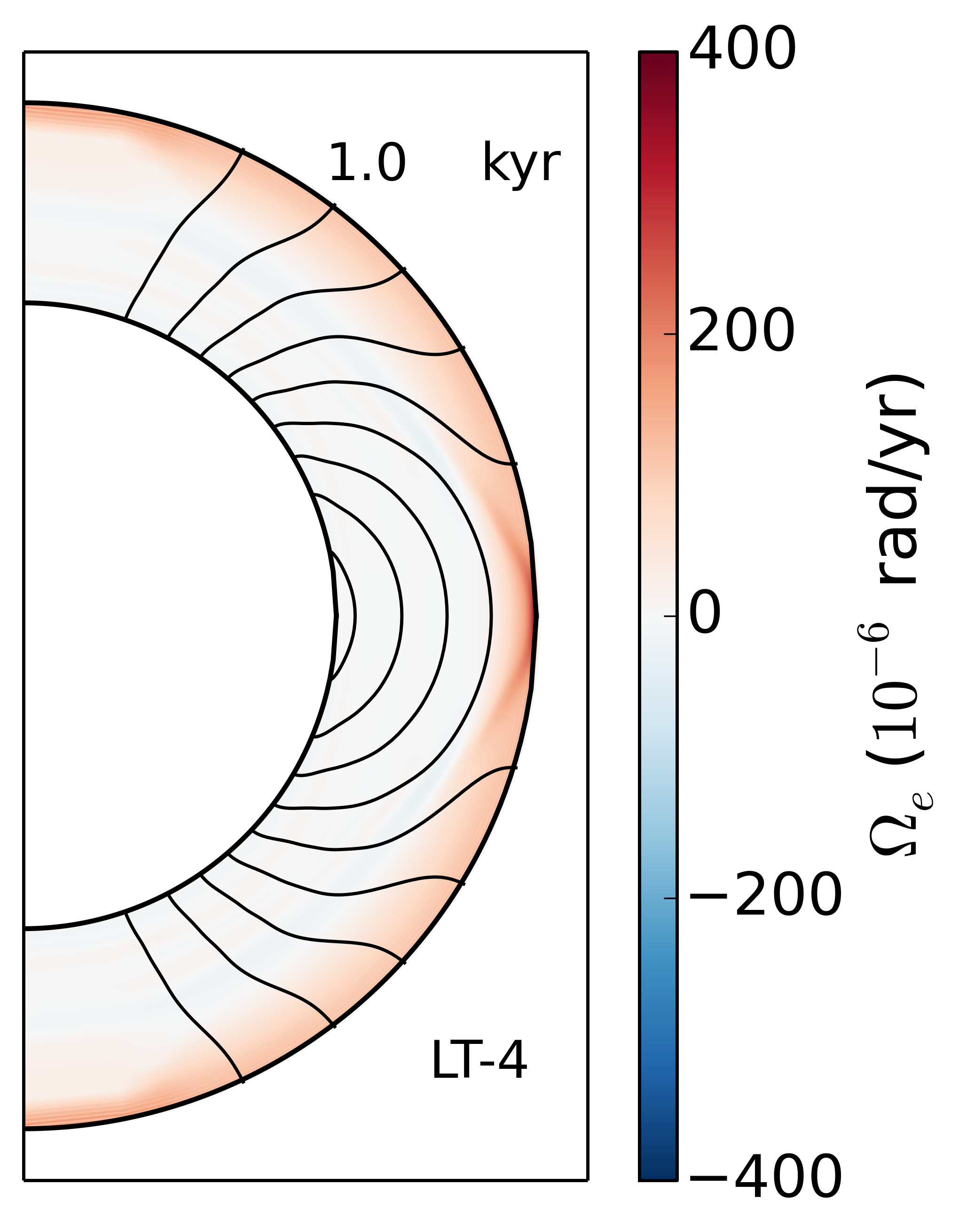}
    \includegraphics[width=0.23\textwidth]{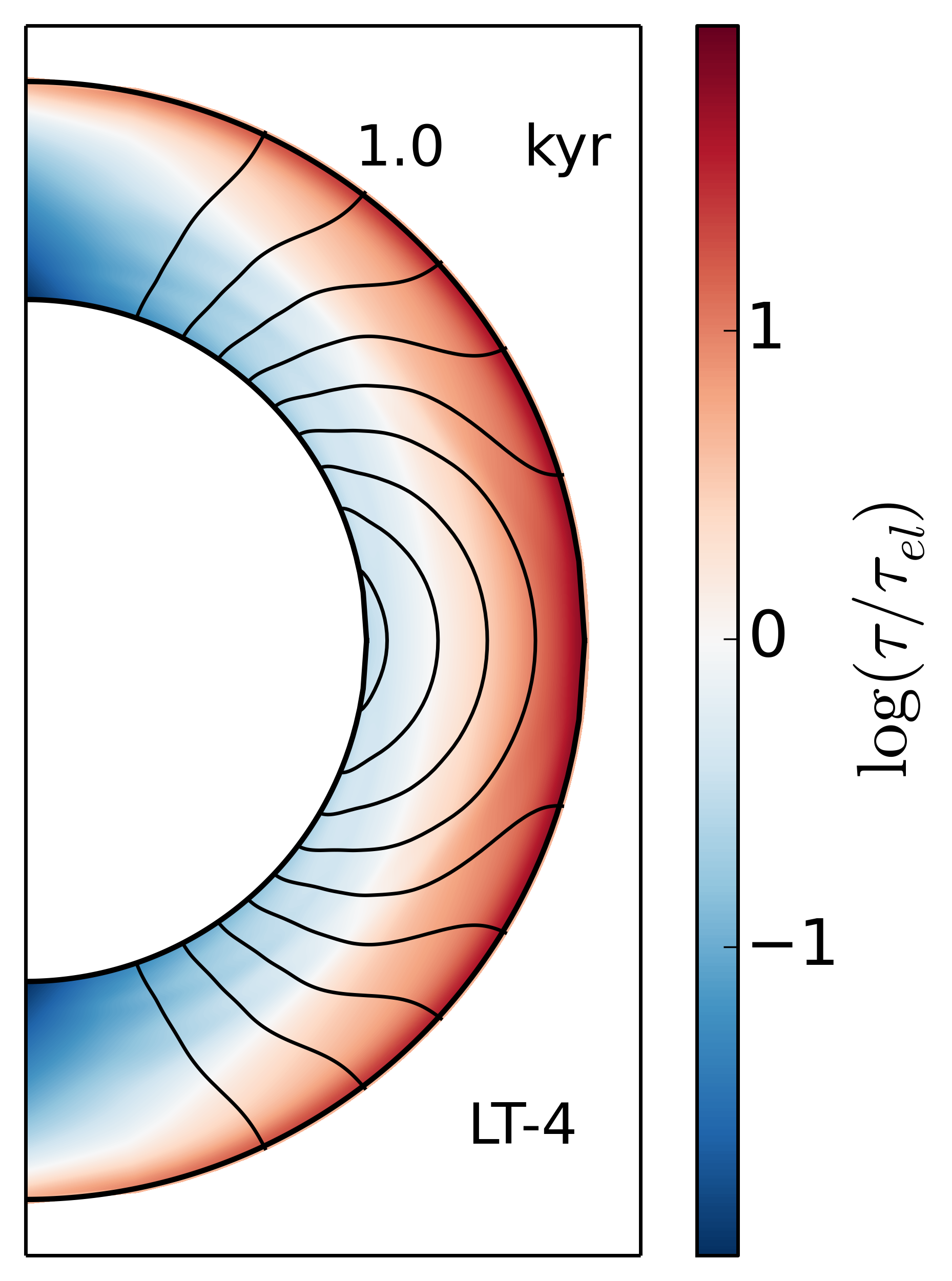}
    \caption{Plots of the electron fluid angular velocity (a) and the logarithm of the ratio of the stress to the critical value (b), in colour, with the poloidal magnetic field lines shown in black, for model LT-4, at 1 kyr.}
    \label{FIG:6}
\end{figure}
%%%%%%%%%%%%%%%%%%%%%%%%%%%%%
\begin{figure}
    \includegraphics[width=0.24\textwidth]{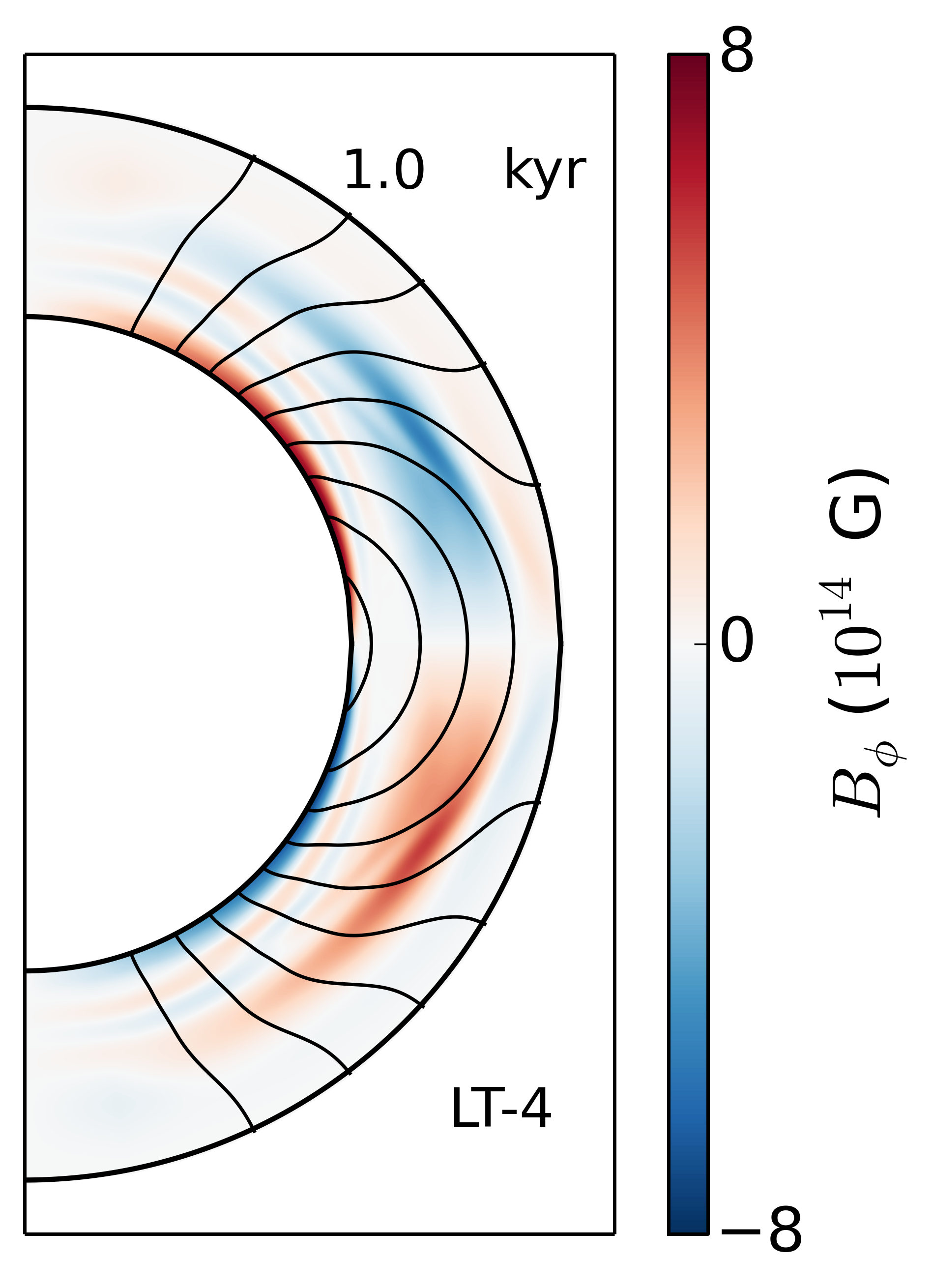}
    \includegraphics[width=0.24\textwidth]{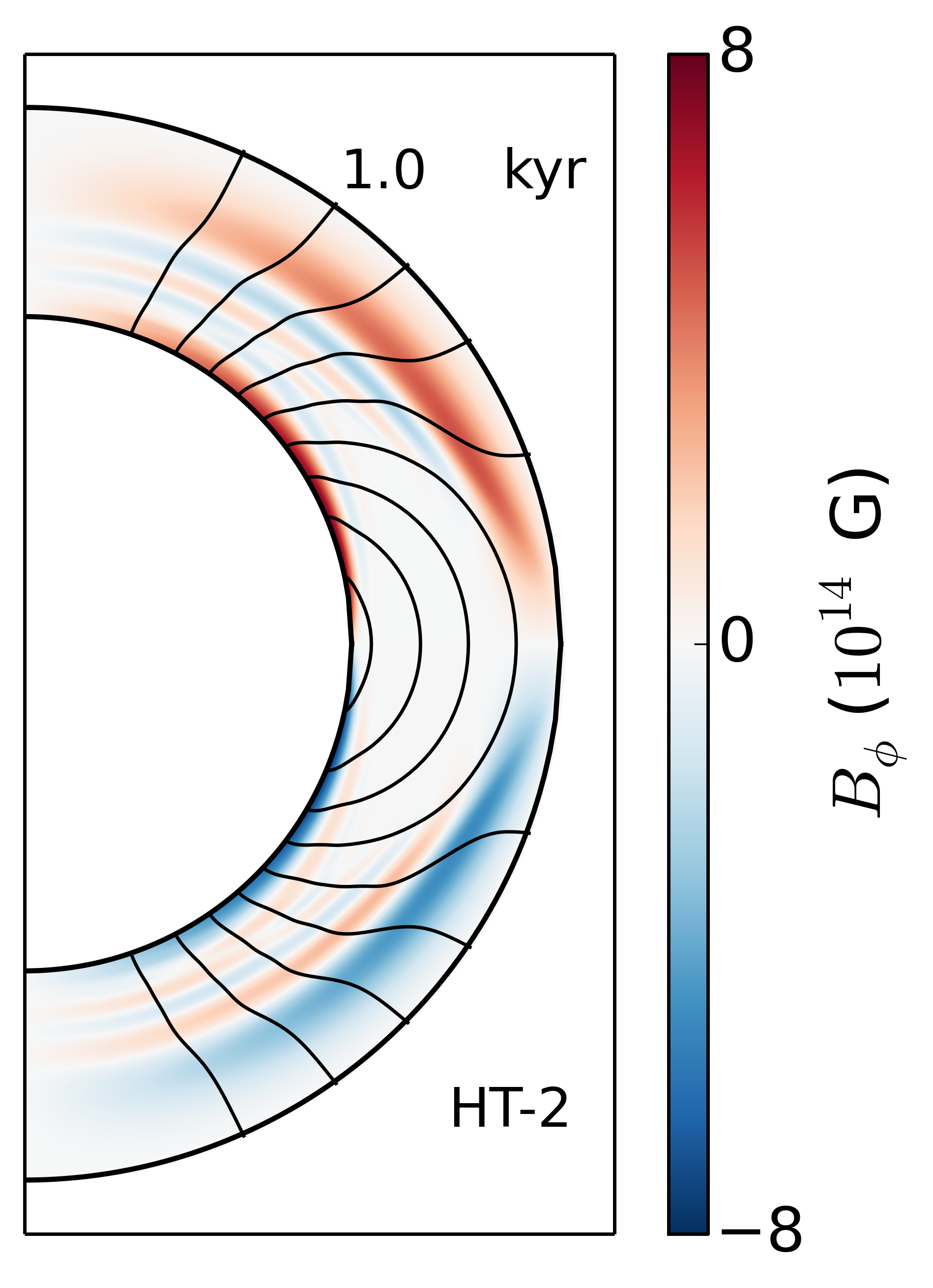}
    \caption{Plots of the toroidal field in colour and the poloidal magnetic field lines in black, for models LT-4 (a) and HT-2 (b) at 1 kyr.}
    \label{FIG:7}
\end{figure}

%%%%%%%%%%%%%%%%%%%%%%%%%%%%%%%%%%%%%%%%%%

We have performed numerical simulations of the magnetic field evolution including plastic flow using the regimes outlined in section \ref{SIMULATIONS}. Given the vast parameter space related to the choice of the initial conditions and the plastic flow viscosity parameter $\nu$, we have focused on the impact of the following choices: whether the plastic flow is local, intermediate or global, the value of $\nu$, and the intensity and structure of the magnetic field -- in particular whether it threads the core or not. The simulation information is presented in Table \ref{TAB:1}. We have adopted the following naming convention. The first letter on the name of the run is either, L, I, G or H depending on the failure being local, intermediate, global or whether there is no failure and the evolution is only due to the Hall effect. The second letter signifies whether the field is confined in the crust (C) or threads the core (T). Finally, to differentiate from simulations that have the same type of failure prescription and structure, but different magnetic field intensity or plastic flow viscosity, we use a different number.

\subsection{Evolution of crust-confined fields}

Let us first consider the simulations where the field is confined in the crust. In all simulations, the stress immediately exceeds the critical value at the equator of the crust and near the north and south poles close to the surface. Following the failure, the subsequent evolution of the plastic velocity profile depends on the type of failure, whether it is L, I and G; one example of each run, after 1 kyr, is shown in Figure \ref{FIG:2}. In the run where the failure is localised in the region where the stress exceeds the critical value (L-type), the overall plastic flow velocity is relatively slow, reaching $5\times 10^{-5}$ rad/yr for model LC-3; Figure \ref{FIG:2} left panel. It becomes somewhat higher for the intermediate case (IC-3) (Figure \ref{FIG:2}, middle panel) and a factor of 4 higher if the plastic flow is allowed to operate everywhere in the crust (model GC-3, Figure \ref{FIG:2} right panel). The plastic flow does not have any noticeable impact on the poloidal field, as can be seen from their similarity in all three panels of Figure \ref{FIG:2}. The angular velocity of the electron fluid depends on the poloidal field structure (see Equation \eqref{Omega}), and so it is also similar for these models.

Given its dependence on the electron number density, the electron angular velocity is highest near the surface of the star; Figure \ref{FIG:3} left panel. The ratio of the stress compared to its critical value is higher near the equator of the star, in the middle of the crust, and near the surface of the star at mid-latitudes; see Figure \ref{FIG:3} right panel. These details depend on the structure of the magnetic field and the changes that occur during its evolution and could vary for a different choice of initial conditions.

While the assumption of failure type has a mild and hardly noticeable impact on the poloidal field, its effect is rather drastic on the toroidal field, as is evident from Figure \ref{FIG:4}. Equation \eqref{dI} shows that the toroidal field is generated from the competition of the electron fluid angular velocity and the plastic flow. Thus, one can see that there is a difference in the intensity of the toroidal field: in the case of the global failure model (GC-3) roughly an order of magnitude weaker than all the others, with its maximum value only  $3\times 10^{13}$G. In this model the plastic velocity efficiently opposes the electron velocity. On the contrary, in the LC-3 and IC-3 models the toroidal field is stronger. In the IC-3 model it is almost annulled in the region where the plastic flow is fastest, but its maximum value is even higher than that of the pure Hall evolution (HC-1). In the LC-3 model the toroidal field has a lower maximum value, but is spread over a larger region of the crust. After 10 kyr, none of the models with plastic flow has such a strong large-scale toroidal field as the pure Hall-Ohmic model. However, the models implementing the local- and intermediate-failure criteria display more complex field geometries than the Hall-Ohmic one, with fields roughly as intense locally as the highest intensity of toroidal field from the Hall-Ohmic model.

The above results refer to the lowest value of plastic flow viscosity we have simulated $\nu_0=10^{37}$g cm$^{-1}$ s$^{-1}$. Simulations with higher values e.g.~$\nu_0=10^{38}$g cm$^{-1}$ s$^{-1}$ have a plastic flow velocity which is lower by a factor of a few. In particular, in the LC-2 run the maximum value of the plastic flow angular velocity is $10^{-5}$rad/yr, and for $\nu_0=10^{39}$g cm$^{-1}$ s$^{-1}$ (LC-1) the maximum plastic flow angular velocity is $10^{-5}$rad/yr. A similar scaling behaviour is observed for intermediate and global failures. In these cases, the toroidal field is more mildly affected by the plastic flow compared to the cases with the lowest $\nu$, and the evolutions become increasingly similar to those without plastic flow; see \citet{Lander:2019} for examples of the effect of increasing $\nu$.

%%%%%%%%%%%%%%%%%%%%%%%%%%%%%%%%%%%%%%%%%%

\subsection{Evolution of core-threading fields}

Next we consider a field that threads the core. These simulations fail near the surface of the star, and generate a flow in the $-\phi$ direction there, which is opposite to the electron fluid velocity. In all three types of failure the flow velocity has the same magnitude, $\sim 10^{-4}$rad/yr, considerably faster than the case of fields confined to the crust alone. This is due to the choice of the initial conditions leading to  faster variation of the poloidal field near the surface in the core-threading fields compared to the crust confined ones. The plastic flow velocity patterns, however, are different deeper in the crust, depending on the type of failure. There is no plastic flow deeper in the crust when the failure is local (LT-4), Figure \ref{FIG:5} left panel. In the intermediate type (IT-4), the plastic flow peaks near the equator and at mid-latitudes and becomes zero near the poles, Figure \ref{FIG:5} middle panel. When the failure is global, the plastic flow velocity peaks near the equator and goes smoothly to zero near the poles. Moreover, in the intermediate and global failure, there is a plastic flow deeper in the crust. The electron fluid angular velocity near the surface of the crust peaks at $4\times 10^{-4}$rad/yr, Figure \ref{FIG:6} left panel. Thus the plastic flow velocity in general opposes the electron flow. Moreover, the ratio of the stress to the critical value exceeds unity in the outer half of the crust at the equator and mid-latitudes, Figure \ref{FIG:6} right panel. This demonstrates why in all models there is a plastic flow near the surface of the star as this region fails for all models. The difference in the plastic flow patterns deeper in the crust reflects the impact of the types of failure studied.

In simulation LT-4 the electron fluid flow in the outer region (left-hand side of Fig. \ref{FIG:6}) and the plastic flow there (left-hand side of Fig. \ref{FIG:5}) have similar magnitude in the outer part of the domain, but opposite direction. Therefore, as for the crustal-confined field in model GC-3 (Fig. \ref{FIG:4}), the Hall source terms from equation \eqref{dI} approximately cancel, and we would expect very limited generation of toroidal field. This is borne out in Fig. \ref{FIG:7}, where we see that the outer region from run LT-4 hosts a far weaker toroidal field than the corresponding model subject to Hall-Ohmic evolution alone (run HT-2).

\subsection{Generic features of evolutions}

Regardless of the details of the field geometry (confined to the crust or otherwise), some aspects of axisymmetric magneto-plastic evolution in a NS crust seem to be universal.

%%%%%%%%%%%%%%%%%%%%%%%%%%%%%%%%%%%%%%%%%%
\begin{figure}
    \includegraphics[width=0.45\textwidth]{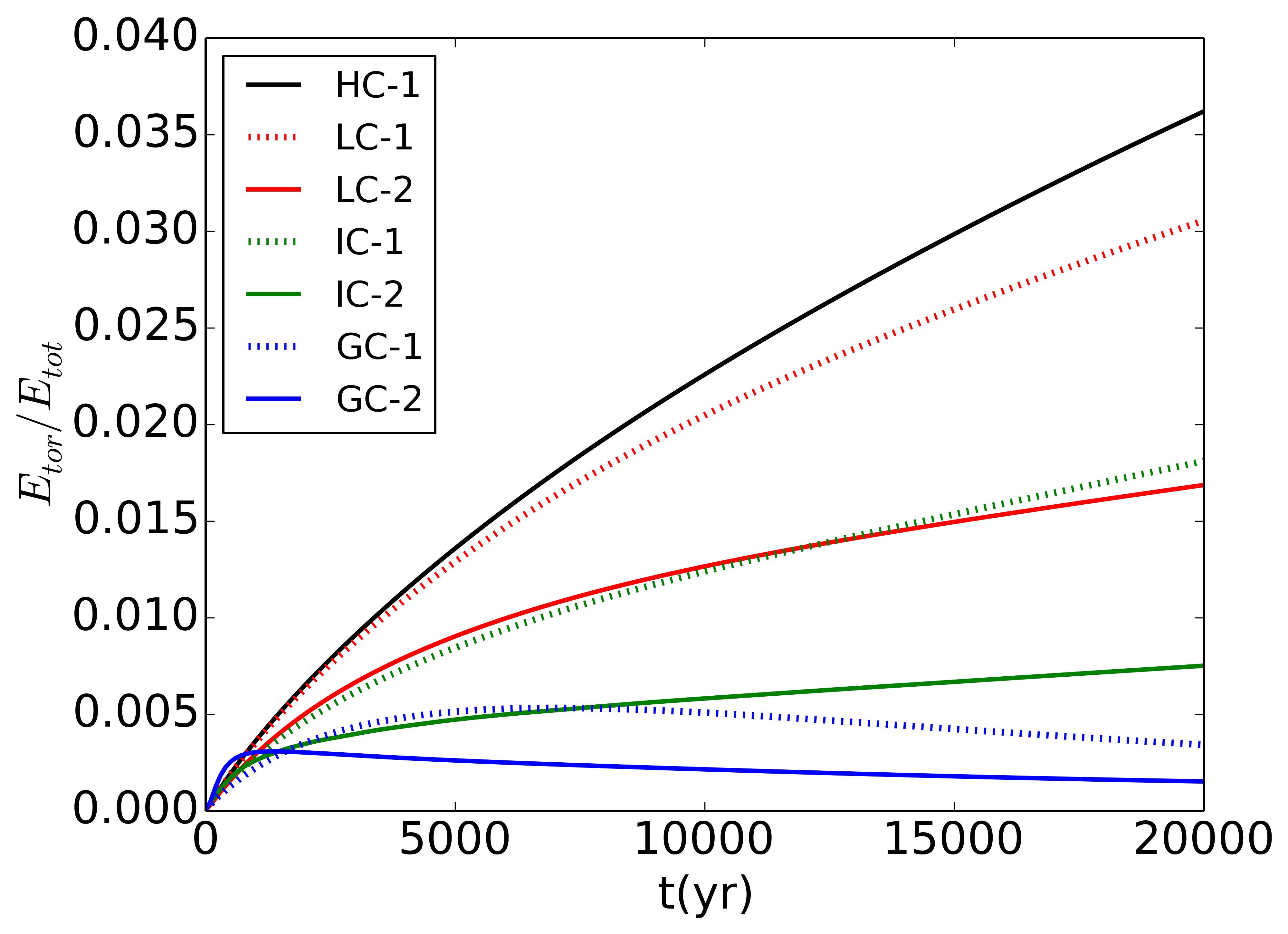}
    \includegraphics[width=0.45\textwidth]{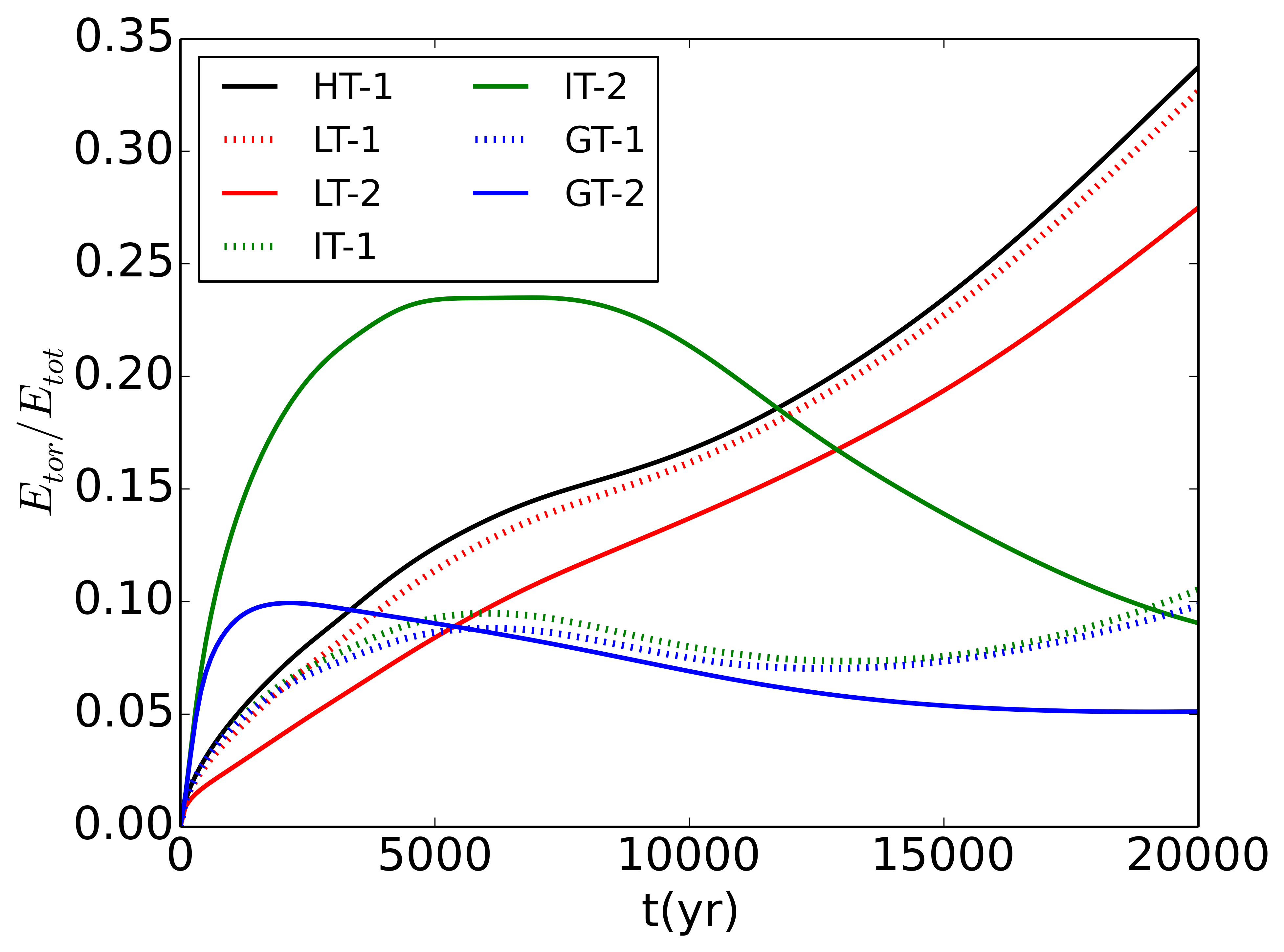}
    \includegraphics[width=0.45\textwidth]{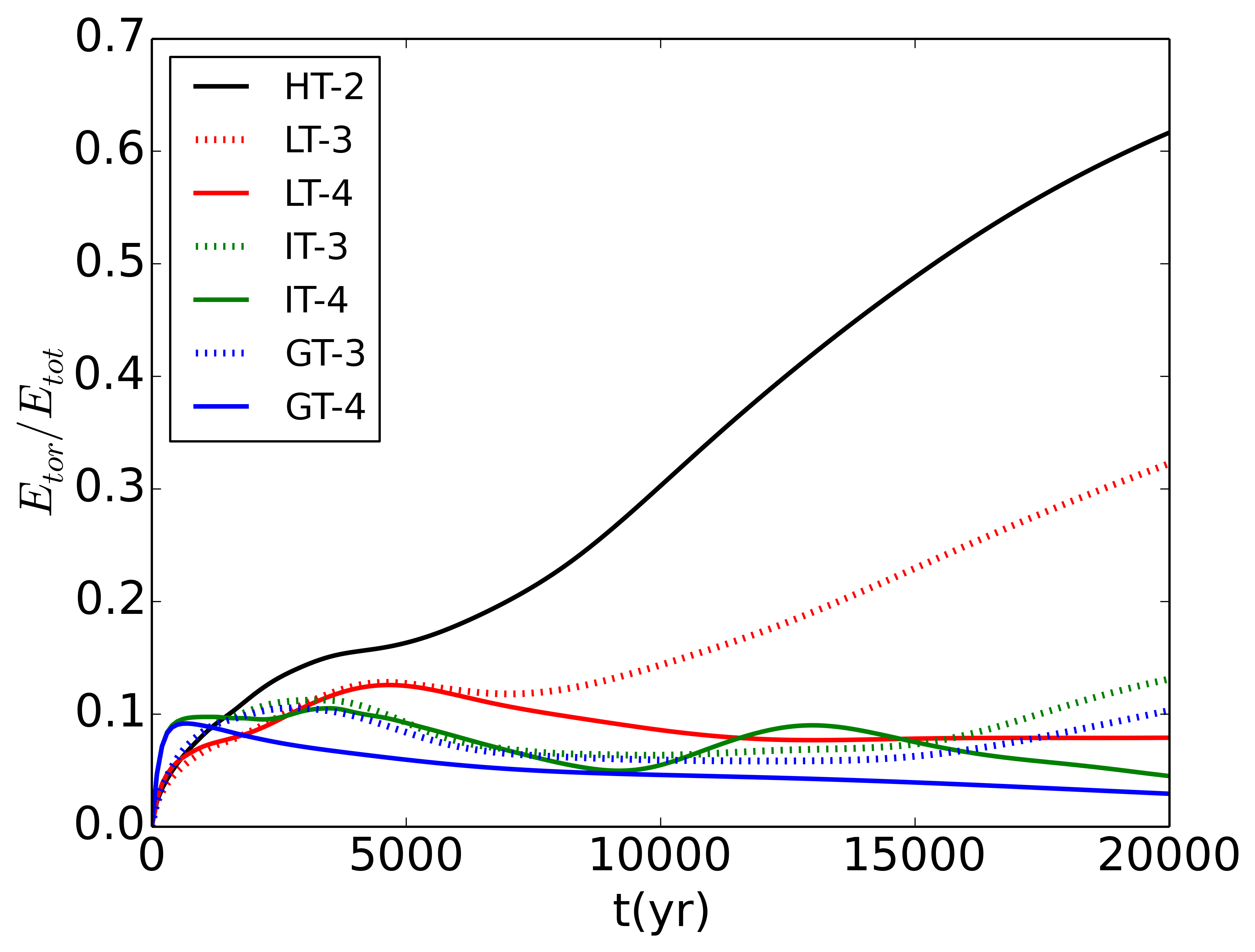}
    \caption{Ratio of the toroidal energy to the total energy for the various runs.}
    \label{FIG:8}
\end{figure}
%%%%%%%%%%%%%%%%%%%%%%%%%%%%%%%%%%%%%%%%%%

Overall, we notice that the formation of the toroidal field is a direct consequence of the differential rotation of the electron fluid, in the context of the Hall effect. Once a plastic flow is present, it has the tendency to counteract this motion. We find this to be the case in the majority of the runs. The effect is more prominent when either the plastic viscosity is lower, or the magnetic field strength higher. When the plastic viscosity is lower, the plastic flow velocity becomes high enough to completely annul the impact of the differential flow of the electron fluid. A stronger magnetic field leads to the failure of a larger fraction of the crust, thus, even in the case of a local failure, the region where the plastic flow occurs is large enough to lead to drastic changes in the toroidal field. 

While the plastic flow has a major impact on the toroidal field, it does not lead to a drastic decrease in the amount of stress in the crust. 
This is mainly due to the fact that the plastic flow impacts the toroidal field directly through Eq. \eqref{dI}, but the poloidal field only indirectly. However, both the poloidal and the toroidal field contribute to the stress in the crust, thus, even if the toroidal field is annulled, any change in the poloidal field will still contribute to the stress.

At later times (t=10kyr), the simulation where the failure is local has a somewhat milder toroidal field near the surface compared to the simulation that evolves only due to the Hall effect. The intermediate and global failures retain their complex plastic flow velocity profile, with the intermediate case alternating from a plastic flow velocity in the $-\phi$ direction to $+\phi$ at mid-latitudes. This has a drastic effect on the toroidal field in the crust changing both its structure and its intensity, compared to systems that evolve only due to the Hall effect. Over very long times, $\gtrsim 10$ kyr, we see from figure \ref{FIG:8} that the fraction of magnetic energy in the toroidal component $E_{tor}/E_{mag}$ is always lower for models with plastic flow than for those evolved with the Hall-Ohmic prescription alone. Furthermore, in evolutions where the field is confined to the crust the maximum value of $E_{tor}/E_{mag}$ is about an order of magnitude smaller than for evolutions of core-threading fields.

\section{Discussion}

\label{DISCUSSION}

The inclusion of a plastic flow in the evolution of the crustal magnetic field adds extra elements of complexity compared to the pure Hall evolution, which is itself a complicated and non-linear problem. Studying an axisymmetric field, intuitively, one would expect that the plastic flow will annul the electron motion in the azimuthal direction. The majority of the simulations have a plastic flow velocity that in general opposes the electron fluid azimuthal velocity, but its profile is  more complicated. For instance, in some simulations it even changes direction, leading to a more drastic twisting of the field compared to pure Hall evolution. This effect is caused because of the following reasons. First, the equation that is solved for the plastic flow velocity is inversely proportional to the plastic viscosity, while the electron velocity depends on the electron number density. While both the plastic viscosity and the electron number density are monotonic functions of the crust density, their functional forms differ drastically. Thus,  even if an extended region of the crust fails and the overall Maxwell stress across this region is similar, the magnitude of the plastic flow velocity can be drastically different due to the large variation of $\nu$ in equation \eqref{Laplacian}. Second, the electron fluid angular velocity depends on $\Psi$ and not $B_{\phi}$. This can be clearly seen from equation \eqref{Omega}, where the electron fluid angular velocity is obtained by the action of the Grad-Shafranov operator (\ref{GSop}) on $\Psi$, without involving $B_{\phi}$ at all. On the contrary, the plastic flow motion depends on $B_{\phi}$ and its derivatives, as it is evident from equation (\ref{Laplacian}). This suggests that while the plastic flow velocity is driven by the magnetic tension term ${\bf B}\cdot \nabla {\bf B}$, the electron fluid angular velocity has a different dependence. A third issue is the dependence of the plastic flow on the type of failure. If the plastic flow is local or intermediate, it will only relieve part of the stress in the regions where the plastic flow is non-zero. Indeed, simulations IC and IT have a toroidal field whose polarity is opposite that of the Hall-only simulations. Similarly, when considering the total energy stored in the toroidal field, we notice that for some models with plastic flow the energy in the toroidal field is temporarily higher than that of the corresponding evolution driven exclusively by the Hall effect, as shown in Figure \ref{FIG:8}. In the long run however, the system that evolves exclusively due to the Hall effect is the one that develops the strongest toroidal field. Nevertheless, this implies that temporarily, the plastic flow is able to twist the field even more drastically and generate a stronger toroidal component. 

The impact of the plastic flow strongly depends on the value of $\nu$, which is largely unknown. Our simulations have allowed us to assess the range of values of $\nu$ for which the plastic flow has a notable effect. In general for $\nu\lesssim 10^{38}$g cm$^{-1}$ s$^{-1}$, the plastic flow velocity is similar to that of the electron fluid. Even if this is the case though, the plastic flow velocity profile is drastically different depending on whether the flow is local or not. A local flow, where only a small fraction of the crust participates and the rest of the star remains intact, does not reach a high velocity. This is mainly due to the boundary conditions imposed, where the rest of the crust does not participate in the flow. Enforcing the flow velocity to be zero beyond the failed region leads to a smaller maximum velocity. On the contrary, if this condition is waived, then the plastic flow velocity can reach much higher values. This effect is very profound if the failed part of the crust is small, but the difference is moderate when the failure affects an extended part of the crust. 

Although the toroidal field is most altered over the course of the evolutions reported here, the poloidal field is affected too, at a somewhat slower rate. These differences are more pronounced at later times ($t>10$kyr). They arise from the term $(\nabla I \times \nabla) 
\cdot \nabla \Psi$ in equation \eqref{dPsi}. As the toroidal field is drastically different this will be reflected in $I$ and consequently in $\Psi$ through this equation. As this is a secondary effect, mediated by this term rather than appearing directly in the field-evolution equation, it is not as profound as the difference in $I$. 

The flow velocity for the runs with $\nu_0 = 2.5\times 10^{38}$g cm$^{-1}$ s$^{-1}$ is in the range of $10^{-5}-10^{-4}$ rad yr$^{-1}$, which in linear velocity  corresponds to $10-100$cm yr$^{-1}$. This velocity is consistent with the findings of our previous work \citep{Lander:2019} and those of \cite{Kojima:2020}. The velocity depends on the intensity of the field as well, with stronger fields leading to higher velocities for given $\nu_0$. We remark however that the presence of a plastic flow does not enhance the decay rate. This might initially seem  counter-intuitive, since one would expect a viscosity term -- like the one in our evolution equations -- to be associated with a dissipative process. However, in our formulation of the problem -- following \cite{Lander:2016}, and based on a simple terrestrial theory of plasticity \citep{Prager:1961} -- $\nu$ simply plays the role of a source term regulating how fast crustal stresses are converted into plastic motion; see equation \eqref{Laplacian} and the discussion in section 2.6 of \cite{Lander:2019}. In fact, in our problem the plastic flow actually reduces the dissipation of magnetic energy, by opposing the formation of the kind of strong currents responsible for the Ohmic decay of the field.

We find that the magnetic-field geometry in a neutron-star crust is sensitive to details of how the crust fails. Depending on whether super-yield stresses cause a failure only in a small region of the crust or a more global collective effect, the magnetic field in the crust evolves in a different way, and in all cases the result is different from the standard Hall-Ohmic (electron MHD) evolution. These differences are not washed out over time, but persist into the crust's old age (for a magnetar) of order 10 kyr. Since high-energy magnetar bursts (and perhaps also fast radio bursts) are often thought to be associated with locally-intense toroidal field (e.g. \citet{Perna:2011}, there is a possibility of using magnetar observations to glean hints of the material physics of the crust. Suppose, for example, that a particular magnetar's activity seems to be associated with emission from its magnetic poles. Then, based on our evolutions from Fig. \ref{FIG:4}, such activity might favour a crust that fails in what we term an `intermediate' manner, since only in this case do we see a high concentration of toroidal field in the polar region alone.

Apart from their bursting activity, magnetars may be displaying the effects of plastic flow in other, less direct, ways. For example, an analysis by \cite{Beloborodov:2016} found mechanical heating due to plastic flow to be one of the most plausible mechanisms for explaining the high surface temperatures of magnetars. Our current model does not include such dissipative effects, but it would be a logical way to extend our work.   Moreover, the impact of the plastic flow could be combined with studies of the magnetothermal evolution that has been  explored in the context of the neutron star diversity \citep{VIGANO:2013, DeGrandis:2020, Igoshev:2021}. 

It has long been known that magnetar rotation and spindown is far noisier and less regular than that of radio pulsars \citep{Melatos:1999,Dib:2008, Tsang:2013}. Their long-term spindown can be irregular, and on shorter timescales they undergo sudden spin-up glitches and potentially (but more controversially) anti-glitches. The key difference between magnetar and pulsar rotation could be that in the former case plastic flow plays a role in the crustal dynamics. Any patch of the inner crust (the region we simulate) will be the pinning site for a number of superfluid neutron vortices. If it begins to move plastically, one of a number of different things could happen to the vortices. 
They could unpin immediately and cause a glitch, or they could remain pinned for some time as the crust shifts and produce a delayed glitch or a more gradual response. If the plastic flow is driven purely by magnetically-induced stresses, it is just as likely to move a patch of crust towards the rotational axis (increasing locally the rotational lag between the superfluid and the rest of the crust) as it is to move it away (decreasing the lag). This could result in a rich phenomenology of timing features; for example, a model for (crustquake-induced) inward vortex motion and its effect on spindown has been explored in relation to the high-magnetic field pulsar J1119-6127 \citep{Antonopoulou:2015,Akbal:2015}.

For now our understanding of magneto-plastic evolution is too rudimentary to make a reliable connection between evolutions and observations, such as the two examples we have outlined above, but we believe that more sophisticated modelling will start to make this a credible possibility.

\section{Conclusions}

\label{CONCLUSIONS}

We have studied the evolution of the magnetic field in the crusts of neutron stars in the presence of a plastic flow. Given that the plasticity of the crust is a largely unknown property, we have explored some regimes that are drastically different from each other, adopting an agnostic view. We find that the evolution of the magnetic field depends on the value of the plastic viscosity and the type of failure we have adopted. 

In general a plastic flow with $\nu_0 =2.5 \times 10^{39}$ g cm$^{-1}$ s$^{-1}$ leads to rather low plastic velocity and has little impact on the magnetic-field evolution. Lower values, of the order $\nu_0 =2.5 \times 10^{38}$ g cm$^{-1}$ s$^{-1}$, lead to flows where the plastic velocity is comparable to the electron fluid velocity, and there is a drastic impact on the field evolution. We remark, though, that the flow does not simply annul the impact of the Hall effect but rather leads to more complex evolution, which depends on the type of failure.

The present study is a step forward from our previous Cartesian geometry work \citep{Lander:2019}. Nevertheless, being confined to an axisymmetric geometry and because of our imposed assumption of incompressibility, the type of flow can only be in the $\phi$ direction. While the lack of flow in the radial direction is dictated by the physics, an axisymmetric flow is a simplification allowing us to tackle a complicated problem. As has already been demonstrated, 3-D studies of the Hall evolution \citep{Gourgouliatos:2016a, Gourgouliatos:2018a} lead to results that are radically and qualitatively different from those found in the axisymmetric problem. We anticipate that a plastic flow in a three dimensional geometry will lead to a more realistic understanding of this effect.

%%%%%%%%%%%%%%%%%%%%%%%%%%%%%%%%%%%%%%%%%%
\section*{Acknowledgments}

We thank Danai Antonopoulou for a helpful discussion about the implications of our results. KNG acknowledges grant FK81641 - Theoretical and Computational Astrophysics.
We thank an anonymous referee for their constructive comments. 

\section*{Data availability statement}
The data underlying this article will be shared on reasonable request to the corresponding author.

%\newpage
\bibliographystyle{mnras}
\bibliography{BibTex.bib}

\begin{thebibliography}{}
\makeatletter
\relax
\def\mn@urlcharsother{\let\do\@makeother \do\$\do\&\do\#\do\^\do\_\do\%\do\~}
\def\mn@doi{\begingroup\mn@urlcharsother \@ifnextchar [ {\mn@doi@}
  {\mn@doi@[]}}
\def\mn@doi@[#1]#2{\def\@tempa{#1}\ifx\@tempa\@empty \href
  {http://dx.doi.org/#2} {doi:#2}\else \href {http://dx.doi.org/#2} {#1}\fi
  \endgroup}
\def\mn@eprint#1#2{\mn@eprint@#1:#2::\@nil}
\def\mn@eprint@arXiv#1{\href {http://arxiv.org/abs/#1} {{\tt arXiv:#1}}}
\def\mn@eprint@dblp#1{\href {http://dblp.uni-trier.de/rec/bibtex/#1.xml}
  {dblp:#1}}
\def\mn@eprint@#1:#2:#3:#4\@nil{\def\@tempa {#1}\def\@tempb {#2}\def\@tempc
  {#3}\ifx \@tempc \@empty \let \@tempc \@tempb \let \@tempb \@tempa \fi \ifx
  \@tempb \@empty \def\@tempb {arXiv}\fi \@ifundefined
  {mn@eprint@\@tempb}{\@tempb:\@tempc}{\expandafter \expandafter \csname
  mn@eprint@\@tempb\endcsname \expandafter{\@tempc}}}

\bibitem[\protect\citeauthoryear{{Akbal}, {G{\"u}gercino{\u{g}}lu},
  {{\c{S}}a{\textcommabelow s}maz Mu{\textcommabelow s}}  \& {Alpar}}{{Akbal}
  et~al.}{2015}]{Akbal:2015}
{Akbal} O.,  {G{\"u}gercino{\u{g}}lu} E.,  {{\c{S}}a{\textcommabelow s}maz
  Mu{\textcommabelow s}} S.,   {Alpar} M.~A.,  2015, \mn@doi [\mnras]
  {10.1093/mnras/stv322}, \href
  {https://ui.adsabs.harvard.edu/abs/2015MNRAS.449..933A} {449, 933}

\bibitem[\protect\citeauthoryear{{Akg{\"u}n}, {Cerd{\'a}-Dur{\'a}n}, {Miralles}
   \& {Pons}}{{Akg{\"u}n} et~al.}{2018}]{Akgun:2018}
{Akg{\"u}n} T.,  {Cerd{\'a}-Dur{\'a}n} P.,  {Miralles} J.~A.,   {Pons} J.~A.,
  2018, \mn@doi [\mnras] {10.1093/mnras/sty2669}, \href
  {https://ui.adsabs.harvard.edu/abs/2018MNRAS.481.5331A} {481, 5331}

\bibitem[\protect\citeauthoryear{{Alford} \& {Halpern}}{{Alford} \&
  {Halpern}}{2016}]{Alford:2016}
{Alford} J.~A.~J.,  {Halpern} J.~P.,  2016, \mn@doi [\apj]
  {10.3847/0004-637X/818/2/122}, \href
  {https://ui.adsabs.harvard.edu/abs/2016ApJ...818..122A} {818, 122}

\bibitem[\protect\citeauthoryear{{Antonopoulou}, {Weltevrede}, {Espinoza},
  {Watts}, {Johnston}, {Shannon}  \& {Kerr}}{{Antonopoulou}
  et~al.}{2015}]{Antonopoulou:2015}
{Antonopoulou} D.,  {Weltevrede} P.,  {Espinoza} C.~M.,  {Watts} A.~L.,
  {Johnston} S.,  {Shannon} R.~M.,   {Kerr} M.,  2015, \mn@doi [\mnras]
  {10.1093/mnras/stu2710}, \href
  {https://ui.adsabs.harvard.edu/abs/2015MNRAS.447.3924A} {447, 3924}

\bibitem[\protect\citeauthoryear{{Beloborodov} \& {Levin}}{{Beloborodov} \&
  {Levin}}{2014}]{Beloborodov:2014}
{Beloborodov} A.~M.,  {Levin} Y.,  2014, \mn@doi [\apjl]
  {10.1088/2041-8205/794/2/L24}, \href
  {https://ui.adsabs.harvard.edu/abs/2014ApJ...794L..24B} {794, L24}

\bibitem[\protect\citeauthoryear{{Beloborodov} \& {Li}}{{Beloborodov} \&
  {Li}}{2016}]{Beloborodov:2016}
{Beloborodov} A.~M.,  {Li} X.,  2016, \mn@doi [\apj]
  {10.3847/1538-4357/833/2/261}, \href
  {https://ui.adsabs.harvard.edu/abs/2016ApJ...833..261B} {833, 261}

\bibitem[\protect\citeauthoryear{{Beloborodov} \& {Thompson}}{{Beloborodov} \&
  {Thompson}}{2007}]{Beloborodov:2007}
{Beloborodov} A.~M.,  {Thompson} C.,  2007, \mn@doi [\apj] {10.1086/508917},
  \href {https://ui.adsabs.harvard.edu/abs/2007ApJ...657..967B} {657, 967}

\bibitem[\protect\citeauthoryear{{Bransgrove}, {Levin}  \&
  {Beloborodov}}{{Bransgrove} et~al.}{2018}]{Bransgrove:2018}
{Bransgrove} A.,  {Levin} Y.,   {Beloborodov} A.,  2018, \mn@doi [\mnras]
  {10.1093/mnras/stx2508}, \href
  {https://ui.adsabs.harvard.edu/abs/2018MNRAS.473.2771B} {473, 2771}

\bibitem[\protect\citeauthoryear{{Chugunov} \& {Horowitz}}{{Chugunov} \&
  {Horowitz}}{2010}]{Chugunov:2010}
{Chugunov} A.~I.,  {Horowitz} C.~J.,  2010, \mn@doi [\mnras]
  {10.1111/j.1745-3933.2010.00903.x}, \href
  {https://ui.adsabs.harvard.edu/abs/2010MNRAS.407L..54C} {407, L54}

\bibitem[\protect\citeauthoryear{{Coti Zelati}, {Rea}, {Pons}, {Campana}  \&
  {Esposito}}{{Coti Zelati} et~al.}{2018}]{CotiZelati:2018}
{Coti Zelati} F.,  {Rea} N.,  {Pons} J.~A.,  {Campana} S.,   {Esposito} P.,
  2018, \mn@doi [\mnras] {10.1093/mnras/stx2679}, \href
  {https://ui.adsabs.harvard.edu/abs/2018MNRAS.474..961C} {474, 961}

\bibitem[\protect\citeauthoryear{Courant, Isaacson  \& Rees}{Courant
  et~al.}{1952}]{Courant:1952}
Courant R.,  Isaacson E.,   Rees M.,  1952, \mn@doi [Communications on Pure and
  Applied Mathematics] {https://doi.org/10.1002/cpa.3160050303}, 5, 243

\bibitem[\protect\citeauthoryear{{Cumming}, {Arras}  \& {Zweibel}}{{Cumming}
  et~al.}{2004}]{Cumming:2004}
{Cumming} A.,  {Arras} P.,   {Zweibel} E.,  2004, \mn@doi [\apj]
  {10.1086/421324}, \href {http://adsabs.harvard.edu/abs/2004ApJ...609..999C}
  {609, 999}

\bibitem[\protect\citeauthoryear{{De Grandis}, {Turolla}, {Wood}, {Zane},
  {Taverna}  \& {Gourgouliatos}}{{De Grandis} et~al.}{2020}]{DeGrandis:2020}
{De Grandis} D.,  {Turolla} R.,  {Wood} T.~S.,  {Zane} S.,  {Taverna} R.,
  {Gourgouliatos} K.~N.,  2020, \mn@doi [\apj] {10.3847/1538-4357/abb6f9},
  \href {https://ui.adsabs.harvard.edu/abs/2020ApJ...903...40D} {903, 40}

\bibitem[\protect\citeauthoryear{{Dib}, {Kaspi}  \& {Gavriil}}{{Dib}
  et~al.}{2008}]{Dib:2008}
{Dib} R.,  {Kaspi} V.~M.,   {Gavriil} F.~P.,  2008, \mn@doi [\apj]
  {10.1086/524653}, \href
  {https://ui.adsabs.harvard.edu/abs/2008ApJ...673.1044D} {673, 1044}

\bibitem[\protect\citeauthoryear{{Douchin} \& {Haensel}}{{Douchin} \&
  {Haensel}}{2001}]{Douchin:2001}
{Douchin} F.,  {Haensel} P.,  2001, \mn@doi [\aap]
  {10.1051/0004-6361:20011402}, \href
  {https://ui.adsabs.harvard.edu/abs/2001A&A...380..151D} {380, 151}

\bibitem[\protect\citeauthoryear{{Esposito}, {Rea}  \& {Israel}}{{Esposito}
  et~al.}{2021}]{Esposito:2021}
{Esposito} P.,  {Rea} N.,   {Israel} G.~L.,  2021, {Magnetars: A Short Review
  and Some Sparse Considerations}.
pp 97--142, \mn@doi{10.1007/978-3-662-62110-3_3}

\bibitem[\protect\citeauthoryear{{Goldreich} \& {Reisenegger}}{{Goldreich} \&
  {Reisenegger}}{1992}]{Goldreich:1992}
{Goldreich} P.,  {Reisenegger} A.,  1992, \mn@doi [\apj] {10.1086/171646},
  \href {http://adsabs.harvard.edu/abs/1992ApJ...395..250G} {395, 250}

\bibitem[\protect\citeauthoryear{{Gourgouliatos} \& {Cumming}}{{Gourgouliatos}
  \& {Cumming}}{2014}]{GOURGOULIATOS:2014a}
{Gourgouliatos} K.~N.,  {Cumming} A.,  2014, \mn@doi [\mnras]
  {10.1093/mnras/stt2300}, \href
  {http://adsabs.harvard.edu/abs/2014MNRAS.438.1618G} {438, 1618}

\bibitem[\protect\citeauthoryear{{Gourgouliatos} \& {Esposito}}{{Gourgouliatos}
  \& {Esposito}}{2018}]{Gourgouliatos:2018b}
{Gourgouliatos} K.~N.,  {Esposito} P.,  2018, {Strongly Magnetized Pulsars:
  Explosive Events and Evolution}.
p.~57, \mn@doi{10.1007/978-3-319-97616-7_2}

\bibitem[\protect\citeauthoryear{{Gourgouliatos} \&
  {Hollerbach}}{{Gourgouliatos} \& {Hollerbach}}{2018}]{Gourgouliatos:2018a}
{Gourgouliatos} K.~N.,  {Hollerbach} R.,  2018, \mn@doi [\apj]
  {10.3847/1538-4357/aa9d93}, \href
  {https://ui.adsabs.harvard.edu/abs/2018ApJ...852...21G} {852, 21}

\bibitem[\protect\citeauthoryear{{Gourgouliatos}, {Wood}  \&
  {Hollerbach}}{{Gourgouliatos} et~al.}{2016}]{Gourgouliatos:2016a}
{Gourgouliatos} K.~N.,  {Wood} T.~S.,   {Hollerbach} R.,  2016, \mn@doi
  [Proceedings of the National Academy of Science] {10.1073/pnas.1522363113},
  \href {http://adsabs.harvard.edu/abs/2016PNAS..113.3944G} {113, 3944}

\bibitem[\protect\citeauthoryear{{Horowitz} \& {Kadau}}{{Horowitz} \&
  {Kadau}}{2009}]{Horowitz:2009}
{Horowitz} C.~J.,  {Kadau} K.,  2009, \mn@doi [Physical Review Letters]
  {10.1103/PhysRevLett.102.191102}, \href
  {http://adsabs.harvard.edu/abs/2009PhRvL.102s1102H} {102, 191102}

\bibitem[\protect\citeauthoryear{{Igoshev}, {Hollerbach}, {Wood}  \&
  {Gourgouliatos}}{{Igoshev} et~al.}{2021}]{Igoshev:2021}
{Igoshev} A.~P.,  {Hollerbach} R.,  {Wood} T.,   {Gourgouliatos} K.~N.,  2021,
  \mn@doi [Nature Astronomy] {10.1038/s41550-020-01220-z}, \href
  {https://ui.adsabs.harvard.edu/abs/2021NatAs...5..145I} {5, 145}

\bibitem[\protect\citeauthoryear{{Karageorgopoulos}, {Gourgouliatos}  \&
  {Contopoulos}}{{Karageorgopoulos} et~al.}{2019}]{Karageorgopoulos:2019}
{Karageorgopoulos} V.,  {Gourgouliatos} K.~N.,   {Contopoulos} I.,  2019,
  \mn@doi [\mnras] {10.1093/mnras/stz1507}, \href
  {https://ui.adsabs.harvard.edu/abs/2019MNRAS.487.3333K} {487, 3333}

\bibitem[\protect\citeauthoryear{{Kaspi} \& {Beloborodov}}{{Kaspi} \&
  {Beloborodov}}{2017}]{Kaspi:2017}
{Kaspi} V.~M.,  {Beloborodov} A.~M.,  2017, \mn@doi [\araa]
  {10.1146/annurev-astro-081915-023329}, \href
  {https://ui.adsabs.harvard.edu/abs/2017ARA&A..55..261K} {55, 261}

\bibitem[\protect\citeauthoryear{{Kojima} \& {Suzuki}}{{Kojima} \&
  {Suzuki}}{2020}]{Kojima:2020}
{Kojima} Y.,  {Suzuki} K.,  2020, \mn@doi [\mnras] {10.1093/mnras/staa1045},
  \href {https://ui.adsabs.harvard.edu/abs/2020MNRAS.494.3790K} {494, 3790}

\bibitem[\protect\citeauthoryear{{Kojima}, {Kisaka}  \& {Fujisawa}}{{Kojima}
  et~al.}{2021}]{Kojima:2021}
{Kojima} Y.,  {Kisaka} S.,   {Fujisawa} K.,  2021, \mn@doi [\mnras]
  {10.1093/mnras/staa3489}, \href
  {https://ui.adsabs.harvard.edu/abs/2021MNRAS.502.2097K} {502, 2097}

\bibitem[\protect\citeauthoryear{{Lander}}{{Lander}}{2013}]{Lander:2013a}
{Lander} S.~K.,  2013, \mn@doi [Physical Review Letters]
  {10.1103/PhysRevLett.110.071101}, \href
  {http://adsabs.harvard.edu/abs/2013PhRvL.110g1101L} {110, 071101}

\bibitem[\protect\citeauthoryear{{Lander}}{{Lander}}{2014}]{Lander:2013b}
{Lander} S.~K.,  2014, \mn@doi [\mnras] {10.1093/mnras/stt1894}, \href
  {https://ui.adsabs.harvard.edu/abs/2014MNRAS.437..424L} {437, 424}

\bibitem[\protect\citeauthoryear{{Lander}}{{Lander}}{2016}]{Lander:2016}
{Lander} S.~K.,  2016, \mn@doi [\apjl] {10.3847/2041-8205/824/2/L21}, \href
  {https://ui.adsabs.harvard.edu/abs/2016ApJ...824L..21L} {824, L21}

\bibitem[\protect\citeauthoryear{{Lander} \& {Gourgouliatos}}{{Lander} \&
  {Gourgouliatos}}{2019}]{Lander:2019}
{Lander} S.~K.,  {Gourgouliatos} K.~N.,  2019, \mn@doi [\mnras]
  {10.1093/mnras/stz1042}, \href
  {https://ui.adsabs.harvard.edu/abs/2019MNRAS.486.4130L} {486, 4130}

\bibitem[\protect\citeauthoryear{{Levin} \& {Lyutikov}}{{Levin} \&
  {Lyutikov}}{2012}]{Levin:2012}
{Levin} Y.,  {Lyutikov} M.,  2012, \mn@doi [\mnras]
  {10.1111/j.1365-2966.2012.22016.x}, \href
  {http://adsabs.harvard.edu/abs/2012MNRAS.427.1574L} {427, 1574}

\bibitem[\protect\citeauthoryear{{Li}, {Levin}  \& {Beloborodov}}{{Li}
  et~al.}{2016}]{Li:2016}
{Li} X.,  {Levin} Y.,   {Beloborodov} A.~M.,  2016, \mn@doi [\apj]
  {10.3847/1538-4357/833/2/189}, \href
  {https://ui.adsabs.harvard.edu/abs/2016ApJ...833..189L} {833, 189}

\bibitem[\protect\citeauthoryear{{Melatos}}{{Melatos}}{1999}]{Melatos:1999}
{Melatos} A.,  1999, \mn@doi [\apjl] {10.1086/312104}, \href
  {https://ui.adsabs.harvard.edu/abs/1999ApJ...519L..77M} {519, L77}

\bibitem[\protect\citeauthoryear{{Passamonti}, {Akg{\"u}n}, {Pons}  \&
  {Miralles}}{{Passamonti} et~al.}{2017a}]{Passamonti:2017a}
{Passamonti} A.,  {Akg{\"u}n} T.,  {Pons} J.~A.,   {Miralles} J.~A.,  2017a,
  \mn@doi [\mnras] {10.1093/mnras/stw2936}, \href
  {https://ui.adsabs.harvard.edu/abs/2017MNRAS.465.3416P} {465, 3416}

\bibitem[\protect\citeauthoryear{{Passamonti}, {Akg{\"u}n}, {Pons}  \&
  {Miralles}}{{Passamonti} et~al.}{2017b}]{Passamonti:2017b}
{Passamonti} A.,  {Akg{\"u}n} T.,  {Pons} J.~A.,   {Miralles} J.~A.,  2017b,
  \mn@doi [\mnras] {10.1093/mnras/stx1192}, \href
  {https://ui.adsabs.harvard.edu/abs/2017MNRAS.469.4979P} {469, 4979}

\bibitem[\protect\citeauthoryear{{Perna} \& {Pons}}{{Perna} \&
  {Pons}}{2011}]{Perna:2011}
{Perna} R.,  {Pons} J.~A.,  2011, \mn@doi [\apjl]
  {10.1088/2041-8205/727/2/L51}, \href
  {http://adsabs.harvard.edu/abs/2011ApJ...727L..51P} {727, L51}

\bibitem[\protect\citeauthoryear{{Pons} \& {Perna}}{{Pons} \&
  {Perna}}{2011}]{Pons:2011}
{Pons} J.~A.,  {Perna} R.,  2011, \mn@doi [\apj] {10.1088/0004-637X/741/2/123},
  \href {http://adsabs.harvard.edu/abs/2011ApJ...741..123P} {741, 123}

\bibitem[\protect\citeauthoryear{{Prager}}{{Prager}}{1961}]{Prager:1961}
{Prager} W.,  1961, {Introduction to Mechanics of Continua}

\bibitem[\protect\citeauthoryear{{Reisenegger}, {Benguria}, {Prieto}, {Araya}
  \& {Lai}}{{Reisenegger} et~al.}{2007}]{Reisenegger:2007}
{Reisenegger} A.,  {Benguria} R.,  {Prieto} J.~P.,  {Araya} P.~A.,   {Lai} D.,
  2007, \mn@doi [\aap] {10.1051/0004-6361:20077874}, \href
  {https://ui.adsabs.harvard.edu/abs/2007A&A...472..233R} {472, 233}

\bibitem[\protect\citeauthoryear{{Thompson} \& {Duncan}}{{Thompson} \&
  {Duncan}}{1995}]{Thompson:1995}
{Thompson} C.,  {Duncan} R.~C.,  1995, \mnras, \href
  {http://adsabs.harvard.edu/abs/1995MNRAS.275..255T} {275, 255}

\bibitem[\protect\citeauthoryear{{Tiengo}, {Esposito}  \&
  {Mereghetti}}{{Tiengo} et~al.}{2008}]{Tiengo:2008}
{Tiengo} A.,  {Esposito} P.,   {Mereghetti} S.,  2008, \mn@doi [\apjl]
  {10.1086/590078}, \href
  {https://ui.adsabs.harvard.edu/abs/2008ApJ...680L.133T} {680, L133}

\bibitem[\protect\citeauthoryear{{Tsang} \& {Gourgouliatos}}{{Tsang} \&
  {Gourgouliatos}}{2013}]{Tsang:2013}
{Tsang} D.,  {Gourgouliatos} K.~N.,  2013, \mn@doi [\apjl]
  {10.1088/2041-8205/773/1/L17}, \href
  {https://ui.adsabs.harvard.edu/abs/2013ApJ...773L..17T} {773, L17}

\bibitem[\protect\citeauthoryear{{Turolla}, {Zane}  \& {Watts}}{{Turolla}
  et~al.}{2015}]{Turolla:2015}
{Turolla} R.,  {Zane} S.,   {Watts} A.~L.,  2015, \mn@doi [Reports on Progress
  in Physics] {10.1088/0034-4885/78/11/116901}, \href
  {https://ui.adsabs.harvard.edu/abs/2015RPPh...78k6901T} {78, 116901}

\bibitem[\protect\citeauthoryear{{Vigan{\`o}}, {Rea}, {Pons}, {Perna},
  {Aguilera}  \& {Miralles}}{{Vigan{\`o}} et~al.}{2013}]{VIGANO:2013}
{Vigan{\`o}} D.,  {Rea} N.,  {Pons} J.~A.,  {Perna} R.,  {Aguilera} D.~N.,
  {Miralles} J.~A.,  2013, \mn@doi [\mnras] {10.1093/mnras/stt1008}, \href
  {http://adsabs.harvard.edu/abs/2013MNRAS.tmp.1674V} {}

\makeatother
\end{thebibliography}

\label{lastpage}

\end{document}